\documentclass[%
 preprint,
 superscriptaddress,
nofootinbib,
 amsmath,amssymb,
 aip,
 jap,
 graphicx
]{revtex4-2}

\usepackage{graphicx}
\usepackage{dcolumn}
\usepackage{tabularx}
\usepackage{booktabs}
\usepackage{bm} 
\usepackage[version=4]{mhchem}
\usepackage[section]{placeins} 
\usepackage{array, makecell} 
\usepackage{siunitx}
\usepackage{pdftexcmds}
\usepackage[journal=rsc]{chemstyle}
\usepackage{microtype}

\usepackage{hyperref}
\usepackage[capitalise]{cleveref} 

\usepackage[myheadings]{fullpage} 
\newcommand{\sub}[1]{$_{\text{#1}}$}
\makeatletter
\newcommand\kv[2]{%
  \ifnum\pdf@strcmp{\unexpanded{#1}}{V}=0 %
     \expandafter\@firstoftwo
  \else
    \expandafter\@secondoftwo
  \fi
    {\textit{#1}\!\sub{#2}}
    {#1\sub{#2}}%
}
\makeatother
\makeatletter
\newcommand\kvc[3]{%
  \ifnum\pdf@strcmp{\unexpanded{#1}}{V}=0 %
     \expandafter\@firstoftwo
  \else
    \expandafter\@secondoftwo
  \fi
    {\textit{#1}\!\sub{#2}$^{#3}$}
    {#1\sub{#2}$^{#3}$}
}
\makeatother

\DeclareSIUnit\angstrom{\protect \text {Å}}

\hypersetup{
    colorlinks=true,                
    breaklinks=true,                
    urlcolor= black,                
    linkcolor= blue,                
    bookmarksopen=false,
    filecolor=black,
    citecolor=magenta,
}

\usepackage{xr}
\makeatletter
\newcommand*{\addFileDependency}[1]{
\typeout{(#1)}
\@addtofilelist{#1}
\IfFileExists{#1}{}{\typeout{No file #1.}}
}\makeatother

\newcommand*{\myexternaldocument}[1]{%
\externaldocument{#1}%
\addFileDependency{#1.tex}%
\addFileDependency{#1.aux}%
}
\myexternaldocument{SI}

\newcommand{\angstrom}{\mbox{\normalfont\AA}} 
\DeclareSIUnit\angstrom{\text {Å}}
\usepackage{booktabs} 
\usepackage{lineno}

\begin{document}

\title{Point defect formation at finite temperatures with machine learning force fields}

\author{Irea Mosquera-Lois}
\affiliation{Thomas Young Centre \& Department of Materials, Imperial College London, London SW7 2AZ, UK}
\author{Johan Klarbring}
\affiliation{Thomas Young Centre \& Department of Materials, Imperial College London, London SW7 2AZ, UK}
\affiliation{Department of Physics, Chemistry and Biology (IFM), Link\"{o}ping University, SE-581 83, Link\"{o}ping, Sweden}
\author{Aron Walsh}
\email{a.walsh@imperial.ac.uk}
\affiliation{Thomas Young Centre \& Department of Materials, Imperial College London, London SW7 2AZ, UK}

\date{\today}
             
\begin{abstract} 
Point defects dictate the properties of many functional materials. The standard approach to modelling the thermodynamics of defects relies on a static description, where the change in Gibbs free energy is approximated by the internal energy. This approach has a low computational cost, but ignores contributions from atomic vibrations and structural configurations that can be accessed at finite temperatures. We train a machine learning force field (MLFF) to explore dynamic defect behaviour using $\mathrm{Te_i^{+1}}$ and \kvc{V}{Te}{+2} in CdTe as exemplars. We consider the different entropic contributions (e.g., electronic, spin, vibrational, orientational, and configurational) and compare methods to compute the defect free energies, ranging from a harmonic treatment to a fully anharmonic approach based on thermodynamic integration. We find that metastable configurations are populated at room temperature and thermal effects increase the predicted concentration of $\mathrm{Te_i^{+1}}$ by two orders of magnitude --- and can thus significantly affect the predicted properties. Overall, our study underscores the importance of finite-temperature effects and the potential of MLFFs to model defect dynamics at both synthesis and device operating temperatures. 
\end{abstract}

\maketitle

\section{Introduction}

Point defects make or break material functionality \cite{stoneham}. They limit photovoltaic efficiency by acting as non-radiative recombination centres
, control ionic conductivity in batteries, provide active sites for catalytic reactions, and platforms for quantum information technologies. Despite their profound effect on the macroscopic properties of crystals, they are present in dilute concentrations and thus render experimental characterisation challenging. As a result, a combination of experiment and theory is needed to understand defect behaviour. 

The key factor when modelling defects is their concentration, which is determined by the free energy of defect formation
, $g_f$, at the synthesis or annealing temperature. Calculating $g_f$ is however computationally challenging, and is thus typically approximated by the formation internal energy
, $u_f(0~\mathrm{K})$, i.e. $g_f(T_{\rm synthesis}) \approx u_f(0~\mathrm{K})$\cite{mosquera-lois_imperfections_2023,freysoldt_first-principles_2014}. Inherent in this approximation is the assumption of a \emph{static} framework, where most studies only consider the defect ground state structure at \SI{0}{K}, and thus neglect metastable configurations that may be populated at the device operating temperature. Since the properties of a defect strongly depend on its geometry\cite{Mosquera-Lois2023,mosquera-lois_search_2021,wang_four-electron_2023,wang_sulfur_2024,wang_upper_2024,squires_oxygen_2024}, the predicted behaviour can be significantly affected when ignoring thermally accessible metastable configurations\cite{yang_non-radiative_2016,alkauskas_role_2016,kavanagh_impact_2022,kavanagh_rapid_2021,kavanagh_intrinsic_2024,guo_nonradiative_2021,fowler_metastable_2024}.

With the development of better computational resources, more accurate studies that go beyond this static \SI{0}{K} approximation are becoming possible using \emph{ab-initio} methods. In the last decades, many investigations have modelled entropic contributions for defects in elementary solids\cite{bochkarev_anharmonic_2019,cheng_computing_2018,chiesa_free_2009,de_koning_atomistic_2002,de_koning_vacancy-formation_2003,glensk_breakdown_2014,grabowski_ab_2009,lucas_vibrational_2009,luo_configurational_2022,mellan_fast_2019,safonova_experimental_2016,satta_vacancy_1998,smirnov_formation_2019,shin_effect_2012,gong_temperature_2018,smirnova_atomistic_2020,zhang_calculating_2018,mathes_breakdown_2020,sinno_atomistic_1996,al-mushadani_free-energy_2003,rauls_entropy_2004,blochl_first-principles_1993,maroudas_calculation_1993,mendelev_molecular_2009,ungar_free_1994,wynblatt_formation_1969,harding_calculation_1985,harding_vibrational_1981,Harding1985,luo_unified_2022,luo_thermodynamic_2022,mishin_calculation_2001,wynblatt_calculation_1969,lapointe_machine_2022,mjacobs_entropy_1990,nam_interpolation_2024,ramos_de_debiaggi_theoretical_2006,carling_vacancy_2003,foiles_evaluation_1994,estreicher_thermodynamics_2004,sanati_first-principles_2003,sanati_defects_2003,sanati_temperature_2005,smargiassi_first-principles_1996,catlow}. 
Thermal effects are harder to model in multinary semiconductors due to the higher number of possible intrinsic charged defects and required level of theory, but have been included for specific defects\cite{agoston_formation_2009,zacherle_ab_2013-1,smith_structural_2023,zacherle_ab_2013,walsh_free_2011,baldassarri_vibrational_2024,millican_redox_2022,moxon_structural_2022,sun_study_2018,miceli_self-compensation_2016,grieshammer_entropies_2013,cazorla_lattice_2017,wynn_structures_2016,youssef_intrinsic_2012,holtzman_equilibrium_2024,gorfer_structure_2024,tarento_comparison_1987,zhang_charge_2024,bjorheim_thermodynamic_2015}. 
However, most of these studies adopt several approximations: i) they only account for vibrational entropies, thereby neglecting other degrees of freedom (e.g., electronic, spin, orientational and configurational), and ii) they adopt the (quasi)harmonic approximation to model vibrational effects (e.g., assuming a quadratic potential energy surface for the interatomic bonds). The limitations of these approximations are system-dependent and not well investigated, and demonstrate the lack of a reliable and affordable approach to model thermal effects for defects. 

In this study, we target these limitations by considering all relevant entropic contributions and systematically comparing the different methods to calculate the defect formation free energy, ranging from a harmonic to a fully anharmonic approach based on thermodynamic integration (TI).
To reduce the cost of these simulations, we use machine learning force fields as a surrogate model, which successfully map the defect energy surfaces (\cref{sec:1}). 
We choose $\mathrm{Te_i^{+1}}$ and \kvc{V}{Te}{+2} in CdTe as exemplar systems since they display potential energy surfaces of different complexity (e.g., presence \emph{versus} lack of low-energy metastable configurations\cite{kavanagh_impact_2022}) (\cref{sec:pes}). 
By modelling the impact of thermal effects on their predicted defect concentrations, we find that these dominate when the defect undergoes symmetry-breaking structural reconstructions and has low-energy metastable configurations, thereby demonstrating the limitations of the idealised \SI{0}{K} description (\cref{sec:conc}). 

\section{Results}

\subsection{Machine learning force fields for defects}\label{sec:1}
Calculating the defect formation free energy, $g_f$, requires modelling the defect formation reaction by computing the free energy difference between products and reactants at the synthesis temperature.\cite{freysoldt_first-principles_2014,mosquera-lois_imperfections_2023} 
For example, the formation of the positively charged tellurium interstitial, $\mathrm{Te_i^{+1}}$, 
can be described by the defect reaction
\begin{equation}
\begin{aligned}
\mathrm{Te}  \xrightarrow{\mathrm{CdTe}} \mathrm{Te_i}^{+1} + e^{-}.
\end{aligned}
\end{equation}
The corresponding formation free energy is defined from the sum of products (charged defect with an electron in the conduction band) minus the sum of reactants (pristine CdTe host and reservoir of Te). 
In the standard first-principles supercell formalism, this formation energy is
given by 
\begin{equation}
\begin{aligned}
\mathrm{g_f = g(Cd_nTe_{n+1}^{+1}) - g(Te) - g([CdTe]_n)} + E_F + E_{corr}
\end{aligned}
\end{equation}
where Te represents the phase that acts as the external source of atoms during synthesis, $E_F$ denotes the Fermi level 
and $E_{corr}$ is the correction energy for charged defects.  
From these terms, the defect free energy, $\mathrm{g(Cd_nTe_{n+1}^{+1})}$,
is the most challenging to compute due to the low symmetry and large supercells required to model defects (e.g., many force calculations in the (quasi)harmonic method). This is exacerbated when going beyond the harmonic approximation since computing the \emph{anharmonic} free energy with TI requires many and long molecular dynamics runs.\cite{calphy,kapil_assessment_2019} 

\begin{figure}[ht]
    \centering
    \includegraphics[width=0.95\linewidth]{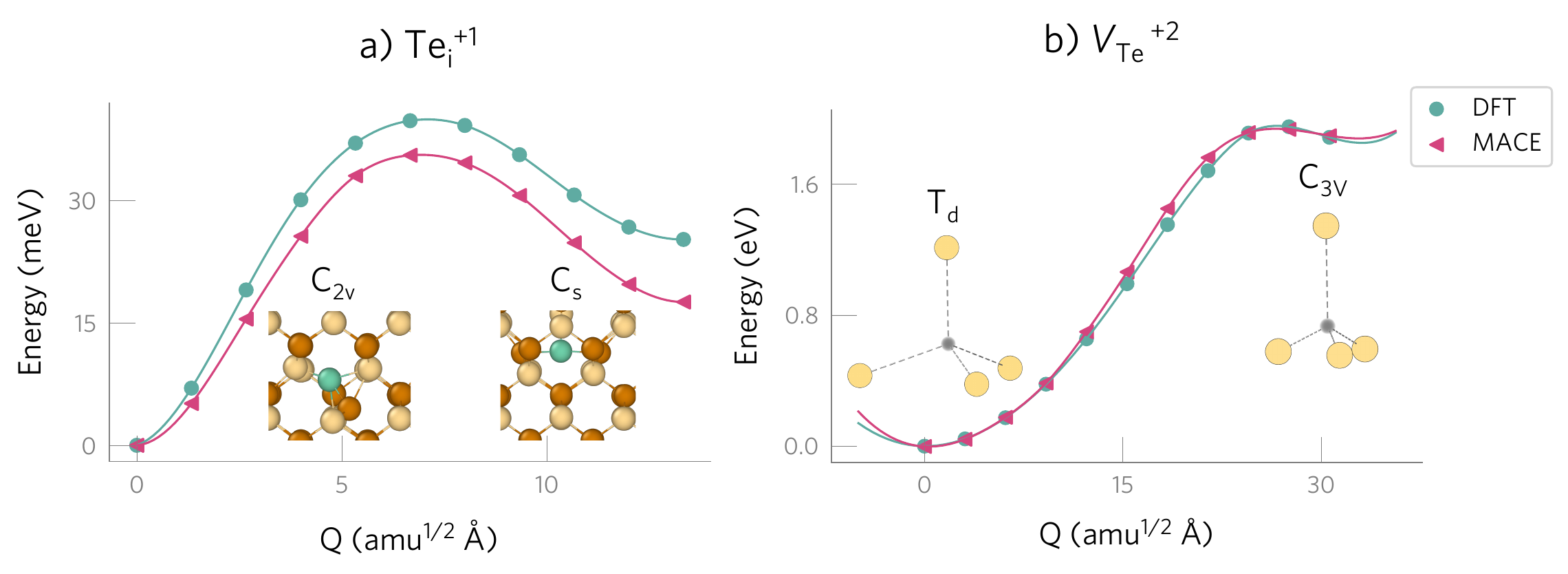}
    \caption{Potential energy surfaces illustrating the defect configurations identified with defect structure searching\cite{shakenbreak2022}, calculated with DFT (green circles) and \texttt{MACE} (pink triangles). a) $\mathrm{Te_i^{+1}}$ is a bistable defect since the metastable configuration (split Te-Cd with $C_{s}$ symmetry) is only 18 meV higher than the ground state (split Te-Te with $C_{\mathrm{2v}}$ symmetry). Note that the differences between DFT and \texttt{MACE} are in meV/\emph{supercell}, with the error in $\Delta E(C_{\mathrm{2v}} - C_{\mathrm{s}})$ only accounting to \SI{0.1}{meV/\emph{atom}}. b) \kvc{V}{Te}{+2}, where the metastable configuration ($C_{\mathrm{3v}}$) is significantly higher in energy (1.8 eV above the ground state structure of $T_d$ symmetry). Te in brown, Cd in yellow, $\mathrm{Te_i^{+1}}$ in green and \kvc{V}{Te}{+2} in shaded grey.}
    \label{fig:pes}
\end{figure}

To reduce the cost of free energy calculations, we employ machine learning force fields and train a separate \texttt{MACE} model\cite{mace} for each system involved in the defect formation reaction, targeting temperatures ranging from \SI{100}{K} to the typical CdTe anneal temperature of \SI{840}{K}\cite{metzger_exceeding_2019}. All models show good accuracies with low mean and root mean square errors on the test set (see \ref{tab:errors} and further discussion in Methods and Supporting Information (SI)). The accuracy of the defect models is further confirmed by mapping the one-dimensional path between the stable defect structures, which shows good agreement despite the small energy difference between the distinct configurations of $\mathrm{Te_i^{+1}}$ (\ref{fig:pes} and SI \ref{sfig:learning_curve_barrier}).

\begin{table}[ht]
\caption{Mean absolute errors 
and root mean square errors (shown in parentheses) 
of the test sets for energies, forces and stresses. The relatively high errors observed for Te are caused by including its liquid phase ($\mathrm{T_{melt}} \approx 704$~K).
Distributions of the absolute errors and the learning curve for the $\mathrm{Te_i^{+1}}$ model are shown in the Supporting Information.}\label{tab:errors}
\vspace{10pt}
\begin{tabular}{cccc}
\hline
System & 
\thead{$\rm Energy$ \\(meV/atom)}  & 
\makecell{$\rm Force$ \\(meV/\AA)} & 
\thead{$\rm Stress$ \\(meV/\AA$^3$)} \\
\hline
CdTe & 0.3 (0.4) & 13 (17)  & 0.2 (0.3) \\
$\mathrm{Te_i^{+1}}$  & 0.5 (0.7)&  21 (30) & 0.2 (0.3) \\ 
\kvc{V}{Te}{+2}  &  0.4 (0.6) & 18 (24) & 0.2 (0.3)  \\
Te  & 1.6 (2.3)&  73 (102) & 0.9 (1.3)   \\
\hline 
\end{tabular}
\end{table}

\subsection{Defect dynamics at room temperature}\label{sec:pes}
We first investigate the limitations of the static framework by comparing the behaviour of the defects at \SI{0}{K} and around the typical operating temperature for a solar cell (\SI{300}{K}). 
The potential energy surface calculated at \SI{0}{K} shows that $\mathrm{Te_i^{+1}}$ is a bistable defect with two accessible structures: a split configuration with either one or two Te-Te bonds\cite{kavanagh_impact_2022}, which have $C_{\mathrm{2v}}$ and $C_{\mathrm{s}}$ site symmetries, respectively, and an energy difference of $\Delta E(C_{\mathrm{s}} - C_{\mathrm{2v}})=18~\mathrm{meV}$ (\ref{fig:pes}). 
In contrast, \kvc{V}{Te}{+2} only has one accessible structure at the device operating conditions since the metastable $C_{3V}$ configuration is 1.8 eV above the $T_{d}$ ground state ($>>k_{B}T=25$ meV at \SI{300}{K}). 

\begin{figure}[ht]
    \centering
    \includegraphics[width=1.0\linewidth]{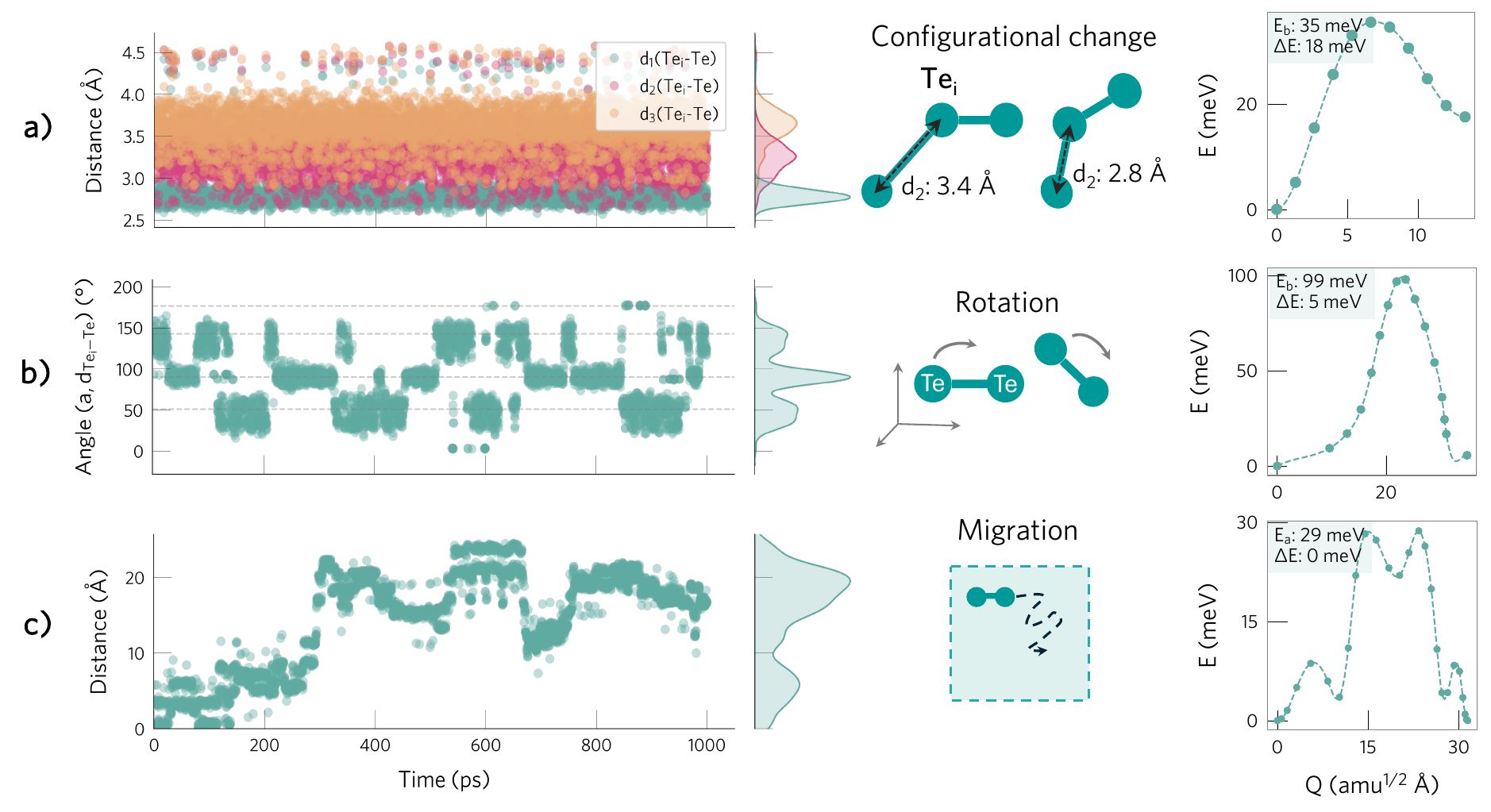}
    \caption{Active degrees of freedom of $\mathrm{Te_i^{+1}}$ at \SI{300}{K} and associated energy barriers. a) Configurational change, reflected by changes in the Te$\mathrm{_i}$-Te distances. The narrow green peak illustrates the shortest Te$\mathrm{_i}$-Te bond, which shows little variation at around 2.8 \angstrom, while the wider pink and orange distributions demonstrate the wide variation in the second and third shortest Te-Te distances (see \ref{sfig:Tei_distance_map}). b) Changes in the orientation of the Te$\mathrm{_i}$-Te  bond with respect to the $\left[100\right]$ direction. c) Migration, illustrated by tracking the distance between the current and original position of the interstitial.}
    \label{fig:dynamics}
\end{figure}

To validate these predictions, 
we perform molecular dynamics under the NPT ensemble (300 K, 1 atm, 1 ns), revealing three distinct motions for $\mathrm{Te_i^{+1}}$ (\ref{fig:dynamics}). The fastest process corresponds to changes in \emph{configuration} between the $C_{\mathrm{2v}}$ and $C_{\mathrm{s}}$ geometries, which is reflected by variations in the distances between $\mathrm{Te_i^{+1}}$ and its neighbouring Te atoms as it alternates between forming 1 and 2 Te-Te bonds. 
On a slower timescale, there are changes in the defect \emph{position} (i.e., hopping between lattice sites) as well as changes in the \emph{orientation} of the Te-Te bond, indicated by variation in the angle between the Te-Te bond(s) and the lattice axes. These three motions occur rapidly on the nanosecond timescale due to their low energy barriers relative to the thermal energy ($E_{b}=28-100$~meV) with rates on the order of $10^{10}~\mathrm{s^{-1}}$ (configurational and hopping) and $10^{8}~\mathrm{s^{-1}}$ (rotation) --- and highlight the configurational, orientational and migration degrees of freedom that contribute to the defect formation entropy.
In contrast, this dynamic behaviour of $\mathrm{Te_i^{+1}}$ differs significantly from that of $\mathrm{V_{Te}^{+2}}$, which remains stable in its $T_d$ ground state configuration, as depicted in \ref{sfig:dynamics_VTe}, and is thus well-described by its static 0 K structure.

\subsection{Impact of defect entropy on predicted concentrations}\label{sec:conc}
The dynamic behaviour of $\mathrm{Te_i^{+1}}$ 
suggests that its formation entropy, $s_f$, will be significant at the CdTe annealing temperature ($\approx 840$~K)\cite{metzger_exceeding_2019} and will thus affect the predicted equilibrium concentration. 
To verify this, we calculate $s_f$ and $g_f$ by considering the different degrees of freedom that change upon forming the defect at a fixed lattice site: electronic, spin, vibrational, orientational and structural (also referred to as configurational).\cite{mosquera-lois_imperfections_2023} 
While the first three terms can be estimated with analytical expressions (see Methods), the vibrational contribution is more challenging and typically requires approximations. 
By assuming a quadratic energy surface for the interatomic bonds, the harmonic vibrational $g_f$ can be calculated, which can be extended to account for thermal expansion with the quasiharmonic approximation\cite{togo_implementation_2023,mosquera-lois_imperfections_2023}. 
However, this harmonic assumption might be limited for defects at high temperatures, where anharmonic effects seem to be important -- as suggested by the high anharmonicity scores\cite{knoop_anharmonicity_2020} observed for $\mathrm{Te_i^{+1}}$ relative to pristine CdTe ($\sigma(840~\mathrm{K})=4.5$ \emph{versus} $0.8$, respectively, and $\sigma$ typically ranging between $0-1$; \ref{sfig:anh_scores}). 

To assess these limitations, we use non-equilibrium thermodynamic integration to account for anharmonic effects and compare the $g_f$ calculated with each approach. 
We do this by starting from the Einstein crystal (independent harmonic oscillators), integrating to the  anharmonic crystal at 100 K, and finally integrating with respect to temperature up to 840 K (see Methods for further details).
Since TI calculates the change in the \emph{total} free energy described by the \texttt{MACE} potential (i.e., the ionic degrees of freedom), it already includes the vibrational, orientational and structural contributions, and thus we only have to add the electronic and spin terms to $g_f^{anh}(T)$. We define three defect formation free energies with increasing accuracy:
\begin{equation}
\begin{aligned}
    g_f^{harm}(T) &= g_f^{vib,harm}(T)  
    - Ts_f^{orient} - Ts_f^{struc}(T) - Ts_f^{spin} - Ts_f^{elec}(T)\\
    g_f^{quasi}(T) &= g_f^{vib,quasi}(T) 
    - Ts_f^{orient} - Ts_f^{struc}(T) - Ts_f^{spin} - Ts_f^{elec}(T)\\
    g_f^{anh}(T) &= g_f^{TI}(T) - Ts_f^{spin} - Ts_f^{elec}(T).
\end{aligned}
\end{equation}
where the (quasi)harmonic vibrational free energy, $g_f^{vib,harm}$, refers to the ground state structure (further details in Methods). 
Here, the (quasi)harmonic approach decouples \emph{all} the degrees of freedom, and thus assumes that the timescales for these processes are sufficiently different to avoid significant mixing\cite{mosquera-lois_imperfections_2023}, while the anharmonic formalism only decouples the electronic from the ionic motions. 

We follow the standard convention in defect chemistry and define $g_f$ as the change in free energy for forming a defect at a \emph{fixed} lattice site (i.e., excluding entropic contributions from the mixing or site entropy\footnote{
This mixing entropy arises from the different ways in which a defect can be arbitrarily placed in the symmetry-equivalent lattice sites and depends on the \emph{equilibrium} defect concentration. Due to this dependence on $c$, it is separated from the free energy of forming the defect at a fixed site ($g_f$, which is independent of $c$ within the dilute limit) when deriving the expression for the defect concentration.\cite{mosquera-lois_imperfections_2023}
}). We calculate the equilibrium defect concentration with
\begin{equation}\label{eq:conc}
[c] = \frac{N_{sites}}{V} \exp{\left(\frac{-g_f}{k_BT}\right)}
\end{equation}
where $V$ denotes the crystallographic unit cell volume and $N_{sites}$ the number of symmetry-equivalent sites where the defect can form in the unit cell.

\begin{figure}
    \centering
    \includegraphics[width=0.95\linewidth]{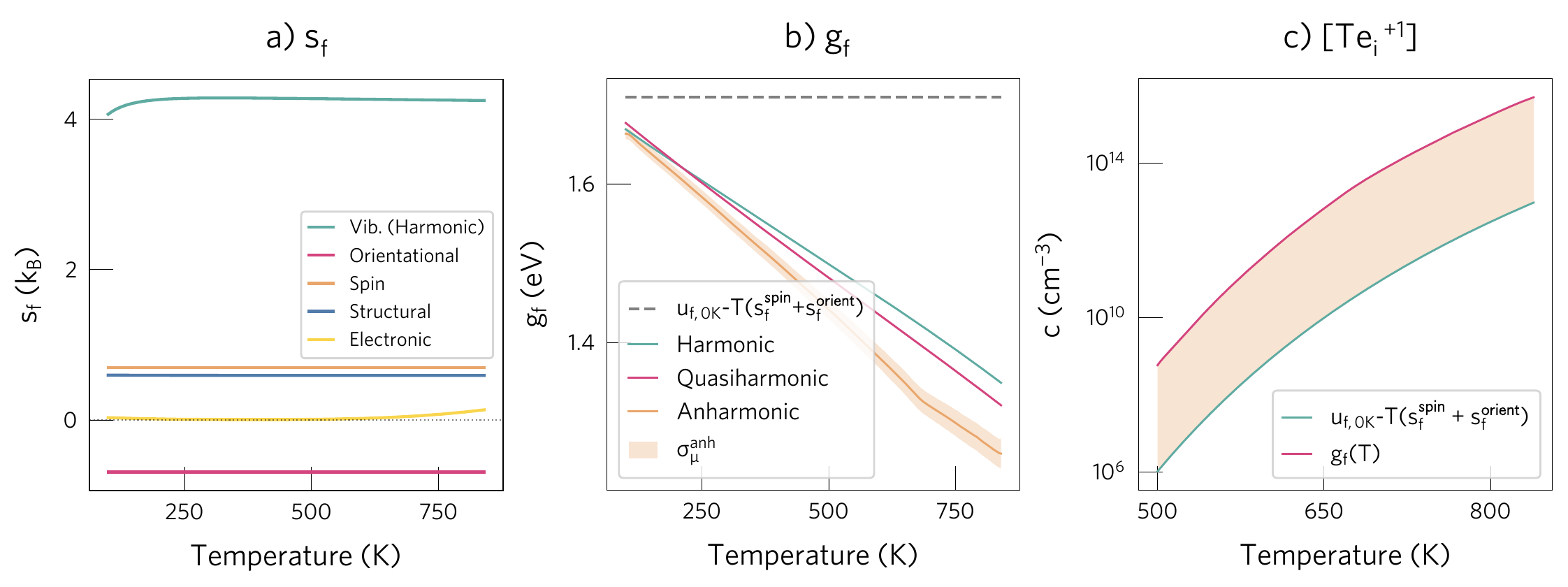}
    \caption{
    a) Contribution of the different degrees of freedom to the formation entropy of $\mathrm{Te_i^{+1}}$. Note that $s_f^{orient}$ is negative since the symmetry increases when going from the initial to the relaxed interstitial structure (see Methods). 
    b) Comparison of approximations for calculating the defect formation free energy of $\mathrm{Te_i^{+1}}$, $g_f(T)$. 
    The shaded orange area illustrates the estimated error in the thermodynamic integration simulations, defined as the standard error of the mean free energy, $\sigma_\mu^{anh}$ (details in Methods). 
    For comparison, the formation internal energy with the spin and orientational entropies, $u_f(0~\mathrm{K})-T(s_f^{spin} + s_f^{orient}$), typically used in most defect studies, is shown with a dashed grey line. 
    c) Effect of including the entropic contribution for predicting the defect concentration.}
    \label{fig:g_f}
\end{figure}

As demonstrated in \ref{fig:g_f}.b, for $\mathrm{Te_i^{+1}}$ thermal effects are significant at annealing temperature, with $g_f(\mathrm{840~K})$ differing by 0.5 eV from $u_f\mathrm{(0~K)}-T(s_f^{spin} + s_f^{orient})$. All methods are in good agreement, indicating that the harmonic approximation gives a reasonable estimate of $g_f$, since 
anharmonic effects approximately cancel out between the bulk and the defect. 
This agreement also validates the decoupling approximation used to separate the different degrees of freedom (e.g., $g_f^{TI} \approx g_f^{vib,harm} - Ts_f^{orient} - Ts_f^{struc}$). 
Using this approximation, we find that their relative entropic contributions 
follow the expected trend, with the vibrational one dominating, followed by the structural, spin, orientational and electronic terms (with $s_f(840~\mathrm{K})$ of 4.2, 0.7, 0.7, -0.7 and 0.1 $k_B$, respectively; \ref{fig:g_f}.a). However, we note that the structural term can become larger for defects that have many low-energy metastable configurations.\cite{luo_configurational_2022} 

Overall, the total entropic contribution is not negligible and significantly affects $g_f$, increasing the predicted concentration by a factor of 500 (\ref{fig:g_f}.c). This importance of entropic effects contrasts with their role in \kvc{V}{Te}{+2}, where they are almost negligible ($g_f(840~\mathrm{K}) - u_f(0~\mathrm{K}) = 0.08 ~\mathrm{eV}$; \ref{sfig:g_f_V_Te}) due to i) smaller magnitude of the vibrational entropy and ii) lack of spin, orientational and structural entropies. As a result, we expect thermal effects to be important for defects which i) introduce strong structural distortions (high $s_f^{vib}$), ii) break the host site symmetry (high $s_f^{orient}$), and iii) have low-energy metastable configurations (high $s_f^{struc}$).


\section{Discussion}
Overall, we have illustrated how to model thermal effects for point defects in crystals, demonstrating the dynamic character of $\mathrm{Te_i^{+1}}$ in CdTe, which rapidly changes between configuration, orientation and position at room temperature. 
These degrees of freedom increase the entropy upon defect formation, and can be computed from standard defect calculations as illustrated in this study. 
The computationally challenging term is the vibrational entropy. We have found that the harmonic approximation gives a reasonable and affordable description of the vibrational formation entropy  at $840~\mathrm{K}$ for a  dynamic defect with a high anharmonicity score. 
This suggests the validity of the harmonic approximation for `simpler' defects with single configurations that do not diffuse in this temperature regime. 

By combining the different entropic contributions, we find that thermal effects increase the predicted concentration of $\mathrm{Te_i^{+1}}$ by two orders of magnitude, and can thus significantly affect the predicted behaviour by shifting the relative defect populations. 
Thermal effects will play a significant role  
for defects that undergo structural reconstructions, break the site symmetry of the host and have low-energy metastable configurations (high $s_f^{vib}$, $s_f^{orient}$ and $s_f^{struc}$), as illustrated by comparing two defects with energy surfaces of different complexity. Beyond defect related factors, hosts with a soft and dynamic lattice or compositional disorder will also be more sensitive to thermal effects, since their defects typically lead to stronger reconstructions (e.g., rebonding or local octahedral rotations in perovskite structures\cite{choi_anti-ferrodistortive-like_2013}) and display many low-energy metastable configurations (e.g., \ce{Sb2Se3}\cite{wang_upper_2024} or alloys\cite{park_accurate_2023} like $\mathrm{CdSe_xTe_{(1-x)}}$\cite{mosquera-lois_machine-learning_2024}). 
A special case is materials where the phase relevant for applications is only stable at finite temperatures. Here, it can be key to model defects at the device operating temperature since their behaviour can be very sensitive to its surrounding structure --- as illustrated by the discrepancies when modelling the carrier capture behaviour of $\mathrm{I_i}$ in different phases of \ce{CsPbI3}\cite{whalley_steric_2023}.

Despite the importance of including thermal effects for accurate defect predictions, the current limitation is the computational cost. While the orientational, spin and electronic terms can be calculated from standard defect calculations using the \texttt{doped} package\cite{doped}, the configurational term requires considering the different thermally-accessible structures of a defect, which can be identified through defect-structure searching methods like \texttt{ShakeNBreak}\cite{shakenbreak2022}. 
More challenging is the vibrational term as it requires going beyond standard static defect calculations. In practice, accounting for this contribution will only be affordable for high-accuracy studies that target the low-energy defects; especially for applications with a high synthesis or operating temperature, like 
industrial thermoelectrics, 
thermochemical water splitting, exhaust automotive catalysts or solid-state fuel cells ($\mathrm{T_{operation}}\approx800-1900$~\si{\kelvin})\cite{naghavi_giant_2017,sikstrom_tutorial_2024,gao_review_2019}. 
In these high temperature applications, thermal effects will populate metastable configurations and could also affect the predicted position of the defect charge transition level (i.e., non-negligible entropic term of $T\times\left[s_f(q,T)-s_f(q',T)\right]/(q-q')$) --- especially when the change in charge state leads to significant differences in the defect structure and symmetry, spin state and position of the defect level within the bandgap.\cite{bjorheim_thermodynamic_2015}   
Finally, we note the promise of machine learning force fields for defects. By learning the defect energy surface, they reduce 
the cost 
of modelling defect dynamics at the device operating temperature and on larger time and length scales, which can be key to predict complex processes like defect reactions or diffusion\cite{zhong_hydrogen_2024}. 

\section{Methodology}

\emph{\textbf{Density Functional Theory calculations.}} All reference calculations were performed with Density Functional Theory using the exchange-correlation functional PBEsol\cite{perdew_restoring_2008} and the projector augmented wave method\cite{Kresse_1996}, as implemented in the Vienna Ab initio Simulation Package (\texttt{VASP})\cite{Kresse_1993,Kresse_1994}. We used the standard PAW PBE potentials (version 64) for Te ($5s^2 5p^4$) and Cd ($4d^{10} 5s^2$). 
Although hybrid functionals are typically required to accurately model the electronic behaviour of defects, we used a more affordable GGA functional for several reasons: i) PBEsol has been found to accurately describe the vibrational properties of crystals\cite{skelton_influence_2015}; 
ii) it correctly identifies the same defect configurations reported in a previous study using the HSE06 hybrid functional\cite{kavanagh_impact_2022}; and 
iii) we aimed to benchmark how to properly train defect MLFFs to reach the high accuracies required to estimate $g_f$, and as a result needed a functional that would allow a thorough exploration of the configurational landscape up to the CdTe synthesis temperature. \\
We converged the plane wave energy cutoff and $\Gamma$-centered k-point mesh to 1 meV/atom, resulting in values of \SI{450}{eV} and $4\times4\times4$ for the conventional cell of CdTe. To minimise Pulay stress errors during molecular dynamics simulations, we increased the converged energy cutoff by 30\% (585 eV). The threshold for electronic convergence was set to $10^{-5}$~eV.

\emph{\textbf{Training of machine learning force fields.}}
We used the structure similarity kernel in \texttt{VASP} to generate the training sets of configurations using its on-the-fly molecular dynamics approach\cite{jinnouchi_phase_2019,jinnouchi_--fly_2019,jinnouchi_--fly_2020,liu_phase_2022}. This involved heating runs performed under the NPT ensemble with a pressure of \SI{1}{atm} and from an initial temperature of \SI{100}{K} up to 30\% above our target temperature of 840 K. In addition, we generated a series of compressed and expanded structures ($0.9-1.1$ of the original cell volume) to ensure that the model could be used for the quasiharmonic approximation. 
For bulk CdTe and its defects, we used a $2\times2\times2$ supercell of the conventional cell (\SI{13.0}{\angstrom} in length and 64 atoms for bulk CdTe). For Te, we included all the low-energy phases available in the Materials Project\cite{jain_commentary_2013} ($E_{hull} \leq k_B T_{synthesis} = k_B \times (840~\mathrm{K})= 0.08~\mathrm{eV}$), which were expanded to cubic supercells of at least \SI{10}{\angstrom} in length. For the liquid Te phase, we generated two models with the \texttt{packmol} code\cite{martinez_p_2009}: two cubic boxes of \SI{15}{\angstrom} and \SI{17.5}{\angstrom} in length, containing 95 and 220 atoms, respectively, which gave densities matching the reported values in previous studies at our target temperatures ($\rho=0.027~\mathrm{atoms/\angstrom^3}$)\cite{akola_density_2010}.

An independent model was trained for each system (bulk CdTe, $\mathrm{Te_i^{+1}}$, \kvc{V}{Te}{+2} and Te) since this lead to higher accuracy models than training one \emph{joint} model on the bulk and defect datasets. 
However, we have observed that training a model \emph{only} on defect configurations should be avoided if the model will be applied to study the absolute energies of larger system sizes (e.g., defect formation energies). While models selectively trained on defect configurations achieve higher accuracy for defective supercells with the \emph{same number of atoms}, these models lead to a \emph{systematic} error in the total energies of \emph{larger supercells} than the supercells used for training, as explained in detail in Supplementary \cref{subsec:mlff_defects}.  
 
After generating the training sets with \texttt{VASP}, we trained  \texttt{MACE}\cite{mace} force fields on these datasets to obtain models with higher accuracy and speed. 10\% of the configurations in these datasets were used as validation sets to monitor the loss during training.
We used a \texttt{MACE} model with Ziegler-Biersack-Littmark (ZBL) pair repulsion\cite{Ziegler1985}, 2 message passing layers, 256 invariant messages, correlation order of 3, angular resolution of 3 and cutoff radius of \SI{5}{\angstrom}. 
The batch size was set to 2 and the Huber loss function was used, with weights of 1, 100 and 100 for the mean square errors in the energies, forces and stresses, respectively. 
For the last 20\% of the training epochs, the weights were updated to values of 1000, 10 and 100 for energy, force and stress, respectively --- following the recommended strategy of increasing the weight on the energy errors during the final training epochs. The models were trained until the validation loss converged, which required around 150-200 epochs. The reference energies were defined as the potential energies of isolated Cd and Te atoms.

\emph{\textbf{Validation of machine learning force fields.}}
To generate the test sets, we performed NPT molecular dynamics simulations with the trained models at three different temperatures (300, 550 and 900 K), running five independent \SI{24}{ps} runs at each temperature. We then sampled 100-300 equally-spaced configurations from these trajectories, and performed DFT calculations on them, which were used to calculate the MAE and RMSE of each model (see distribution of sampled configurations and associated errors in \cref{sfig:errors_CdTe,sfig:errors_Tei,sfig:errors_Te,sfig:errors_VTe,sfig:dataset_Tei}). 
The MAE and RMSE for the forces and stresses were calculated component wise, as defined in Ref.\citenum{morrow_how_2023}. In addition, we also validated that the model successfully described the phonons and vibrational free energy of bulk CdTe (see \ref{sfig:CdTe_phonon_dispersion}). 

\emph{\textbf{Defect calculations.}} Defect calculations were setup and analysed using \texttt{doped}\cite{doped}. To account for spurious finite-size supercell effects, the Kumagai-Oba\cite{kumagai_electrostatics-based_2014} (eFNV) charge correction scheme was used to calculate $E_{corr}$, as automated in \texttt{doped}. The Fermi level was assumed to be located in the middle of the band gap. 

\emph{\textbf{Spin degeneracy.}} The spin degeneracy was calculated with \texttt{doped} using
\begin{equation}
    \Omega^{\mathrm{spin}} = \frac{Z_d^\mathrm{spin}}{Z_b^\mathrm{spin}}=2S+1 
\end{equation}
where $Z$ denotes the partition function and $S$ the total spin angular momentum. For example,  $\mathrm{Te_i^{+1}}$ has one unpaired electron, resulting in $\Omega^{\mathrm{spin}}=2\times(1/2)+1=2$. This degeneracy factor $\Omega$ can be converted into its respective formation entropy using 
$s_f = k_B \ln{(\Omega)}$\cite{mosquera-lois_imperfections_2023}, where $k_B$ denotes the Boltzmann constant.

\emph{\textbf{Orientational degeneracy.}} The orientational degeneracy was also calculated with \texttt{doped} using
\begin{equation}
\Omega^{\mathrm{orient}} = \frac{Z^{\mathrm{orient}}_d}{Z^{\mathrm{orient}}_b} = \frac{N_b}{N_d}
\end{equation}
where $N$ is the number of symmetry operations of the defect site in the bulk ($b$) and defective ($d$) supercells\cite{mosquera-lois_imperfections_2023}. As discussed in the \texttt{doped} documentation\cite{doped}, for vacancies and substitutions there is a clear definition of the defect site in the pristine supercell (e.g., the lattice site where the vacancy/substitution forms). In contrast, for interstitials, the definition of the lattice site in the bulk can be ambiguous, which affects the \emph{partition} between orientational degeneracy and site multiplicity. Here, we follow the definition adopted by Kavanagh \emph{et al}\cite{doped}, where the interstitial site in the \emph{bulk} is defined as the \emph{relaxed} site of the interstitial but with all other atoms fixed in their \emph{bulk} (unrelaxed) positions, while the interstitial site in the \emph{defect} supercell corresponds to the relaxed position of \emph{both} the interstitial and all other atoms. Accordingly, the site multiplicity is determined for the lattice site that the interstitial occupies after relaxation and with all other atoms fixed in their bulk positions. Note that other definitions can be adopted and will lead to the \emph{same total} prefactor ($\Omega^{\mathrm{orient}} \times N_{site}$) but different \emph{partitions} into the orientational and site degeneracies. 

For $\mathrm{Te_i^{+1}}$, the
$C_{\mathrm{2v}}$ configuration has an initial site symmetry of $C_{\mathrm{s}}$ which becomes $C_{\mathrm{2v}}$ when the atoms around the interstitial relax due to the formation of the Te-Te dimer. Similarly, the (metastable) $C_{\mathrm{s}}$ configuration has an initial $C_1$ site symmetry which becomes $C_{\mathrm{s}}$ after the relaxation. As a result, for both configurations the orientational degeneracy $\Omega^{\mathrm{orient}}$ is $0.5$ (e.g. the site symmetry increases upon relaxation of the atoms around the interstitial). The site multiplicities per primitive cell are 12 and 24 for the $C_{\mathrm{2v}}$ and $C_{\mathrm{s}}$ configurations, respectively. These degeneracies are accounted for when predicting the defect concentration (\cref{eq:conc}) where the orientational entropy is included in $g_f$.
%

\begin{sloppypar}
\emph{\textbf{Electronic entropy.}} The electronic entropy was calculated using the fixed density of states (DOS) approximation\cite{Eriksson1992,zhang_accurate_2017,metsue_contribution_2014,willaime_electronic_2000,estreicher_thermodynamics_2004}, that assumes a temperature-independent DOS. Since the electronic entropy is sensitive to the bandgap, and PBEsol significantly underestimates it, we performed self-consistent field HSE06 ($\alpha=0.345$\cite{kavanagh_rapid_2021}) calculations on the 
 \SI{0}{K} structures optimised with PBEsol. The electronic entropy is then calculated using
\end{sloppypar}
\begin{equation}\label{eq:elec_entropy}
    S^\mathrm{elec} = 
    - \gamma k_B 
    \int_{-\infty}^{\infty} D(E) 
    \left(
        f(E, T) 
        \ln{\left( f(E, T) \right)}
        + 
        \left( 1 - f(E, T) \right) 
        \ln{\left( 1 - f(E, T) \right)}
    \right)
    dE
\end{equation}
where $\gamma$ equals 1 for spin-polarized systems and 2 for spin-unpolarized systems.\cite{zhang_accurate_2017} $D(E)$ is the electronic density of states at energy $E$ (calculated at \SI{0}{K}) and $f(E)$ is the occupation of the energy level $E$ given by Fermi-Dirac occupation statistics
\begin{equation}
    f (E, T) = \left( 
    \exp{\left( \frac{E - E_F}{k_B T}\right)} + 1 
    \right) ^{-1}
\end{equation}
with $E_F$ denoting the Fermi level. 
We define the formation electronic entropy as the entropy change in the reaction $\mathrm{(CdTe)_{32} + Te \rightarrow Cd_{32}Te_{33}^{+1} + e^{-}}$ (\ref{sfig:s_elec}), with $E_F$ for each of these terms calculated as follows:
\begin{itemize}
    \item Te, CdTe and $\mathrm{Cd_{32}Te_{33}^{+1}}$: The Fermi level is assumed to be located mid-gap between the highest occupied and lowest unoccupied state. Note that, in theory, the Fermi level of CdTe should correspond to the \emph{self-consistent} value determined for a set of defects and charge states\cite{squires_py-sc-fermi_2023,buckeridge_equilibrium_2019}, but in practice this is currently not feasible, and assuming the level to be midgap is a reasonable approximation in this study\cite{blochl_first-principles_1993}. 
    $\mathrm{Te_i^{+1}}$ introduces an empty state 0.75 eV above the valence band maximum (VBM), and thus $E_F$ is set to $0.75/2=0.37$ eV above the VBM. Note that the electronic entropy is sensitive to the 
    energy difference between $E_F$ and the lowest unoccupied state, thus requiring an accurate electronic structure.
    \item Extra electron: It is defined as the excess electronic entropy when one extra electron is added to the conduction band minimum (CBM) of the bulk system, i.e. $s^{elec}(N+1) - s^{elec}(N)$, where $s^{elec}(N)$ denotes the electronic entropy of bulk CdTe with $N$ and $N+1$ electrons. To calculate the electronic entropy of the $N+1$ system, we determine $E_F$ with\cite{smets_solar_2016} 
    \begin{equation}
        E_F = E_{CBM} + k_B T \ln\left(\frac{n}{N_C}\right)
    \end{equation}
    where $N_C$ denotes the effective density of states in the conduction band, given by 
    \begin{equation}
    N_c = 2 \left( \frac{2 \pi m_e^* k_B T}{h^2} \right)^{3/2}
    \end{equation}
    with $m_e^*$ and $h$ representing the electron effective mass and Planck constant, respectively. For the concentration of excess electrons donated by the defect, $n$, we assume a value of $n=10^{15}$~cm$^{-3}$ in the dilute limit of defect formation. 
\end{itemize}
Finally, we note that the electronic entropy can become significant at elevated temperatures ($T\geq1000$~K, \ref{sfig:s_elec}) if the defect i) introduces an (occupied) empty state close to the (CBM) VBM or ii) changes the occupation of localised d/f bands of nearby cations (e.g., \kvc{V}{O}{+2} reducing two $\mathrm{Ce^{+4}}$ to $\mathrm{Ce^{+3}}$ in \ce{CeO2}), as demonstrated in previous studies.\cite{naghavi_giant_2017,lany_communication_2018}

\emph{\textbf{Structural entropy.}} The code \texttt{ShakeNBreak} was used to identify the defect ground state and metastable configurations.\cite{shakenbreak2022,Mosquera-Lois2023,mosquera-lois_search_2021} From these configurations, the structural or configurational entropy can be estimated using 
\begin{equation}\label{eq:s_struc}
s_f^{struc} = -k_B\sum_i{p_i\ln(p_i)}\\
\end{equation}
where $p_i$ denotes the Boltzmann probability of configuration $i$, given by
\begin{equation}\label{eq:prob}
p_i = \frac{e^{-\frac{G_i}{k_BT}}}{\sum_{j}{e^{-\frac{G_j}{k_BT}}}} = \frac{\Omega_i^{elec}\Omega_i^{spin}\Omega_i^{orient}\Omega_i^{vib}e^{-\frac{U_i}{k_BT}}}{\sum_{j}{\Omega_j^{elec}\Omega_j^{spin}\Omega_j^{orient}\Omega_j^{vib}e^{-\frac{U_j}{k_BT}}}}
\end{equation}
where $G_i$ and $U_i$ are the Gibbs free energy and internal energy of configuration $i$, and we have included the degeneracy factors $\Omega$ since these can be configuration dependent. In practice, the main degrees of freedom 
that change between configurations are the orientational and spin (and vibrational to a lower extent), thus simplifying \cref{eq:prob} to
\begin{equation}
p_i = \frac{\Omega_i^{spin}\Omega_i^{orient}e^{-\frac{U_i-TS_i^{vib}}{k_BT}}}{\sum_{j}{\Omega_j^{spin}\Omega_j^{orient}e^{-\frac{U_j-TS_j^{vib}}{k_BT}}}}.
\end{equation}
%
%
%
Beyond applying this analytical approach, we also calculated the structural entropy using the `inherent structures' method (IS).\cite{luo_configurational_2022} Within this formalism, we performed NPT MD trajectories at 30 temperatures (equally spaced ranging from 100 to \SI{840}{K}), then sampled 1600 equally-spaced configurations and performed conjugate gradient optimisations to quench the structures to the nearest local minima in the \SI{0}{K} PES (\ref{sfig:ratio_configs_Tei}). The configurational entropy was then calculated with
\begin{equation}
s_f^{struc} = s_f^{struc}(T_i) + \int_{T_i}^{T_f}{\frac{1}{T} \frac{\partial \left< e_{IS}^{V}(T) \right> }{\partial T} dT}.
\end{equation}
where $\left< e_{IS}^{V}(T) \right>$ denotes the average potential energy of the inherent structures sampled at $T$, with the volume fixed to the optimal value for the \SI{0}{K} ground state structure, and $T_i$ and $T_f$ set to 100 and \SI{840}{K}, respectively. This method resulted in $s_f^{struc}(840~\mathrm{K})=1.05~k_B$, in the same order of the value of $0.6~k_B$ obtained with the analytical method. The slightly larger value of $s_f^{struc}$ estimated with the IS formalism likely stems from the additional intermediate configurations sampled with quenching.

Finally, a third approach to account for the structural entropy involves coarse-graining the configurational degree of freedom, and calculating the total defect concentration as a sum over the different configurations $i$ using\cite{mosquera-lois_imperfections_2023}
\begin{equation}
    c = \sum_i c_i = \sum_i \frac{N_i}{V} \exp{\left(-\frac{g_{f,i}}{{k_B T}}\right)}
\end{equation}
where $N_i$ and $g_{f,i}$ denote the number of symmetry-equivalent sites and formation free energy of configuration $i$ (with $g_{f,i} = u_{f,i} - T (s^{vib,harm}_{f,i} + s^{orient}_{f,i} + s^{spin}_{f,i} + s^{elec}_{f,i})$). From this expression, an effective formation free energy can be obtained by 
\begin{equation}
    g_{eff} = - k_B T \ln{\left(\frac{V}{N} c \right)}
\end{equation}
where $N$ denotes the number of symmetry-equivalent sites for the ground state structure. By comparing $g_{eff}$ with the $g_f$ calculated using \cref{eq:s_struc}, we verified that these values agree (see \ref{sfig:s_struc_coarse_grained}).

\emph{\textbf{Vibrational entropy.}} The harmonic and quasiharmonic vibrational free energies were calculated using \texttt{phonopy}\cite{togo_first_2015,togo_implementation_2023}. Within the quasiharmonic framework, which includes the effect of thermal expansion on the phonon frequencies, we generated 11 structures by scaling the supercell volume by factors ranging from 0.9 to 1.10 with a 0.02 increment. 

\emph{\textbf{Thermodynamic integration.}}
Fully anharmonic free energies were calculated using non-equilibrium TI in \texttt{LAMMPS}\cite{LAMMPS,freitas_nonequilibrium_2016}, as implemented in the code \texttt{calphy}\cite{calphy}. This involved two thermodynamic integration paths: first, we integrate from the Einstein crystal to the anharmonic one at \SI{100}{K} (Frenkel Ladd method) and then we calculate the temperature variation of the free energy at constant pressure by integrating from \SI{100}{K} to \SI{840}{K} (commonly known as reversible scaling). These simulations were performed fixing the center of mass, as done in previous studies\cite{cheng_computing_2018,calphy,freitas_nonequilibrium_2016}. 
For each path, we performed 10 independent TI runs to estimate the error, defined as the standard error of the mean free energy ($\sigma_\mu = \sigma / \sqrt{N}$, where $\sigma$ denotes the standard deviation in the individual free energies calculated from the $N=10$ independent TI runs). The switching time was adjusted for each system until the error converged to an acceptable value (see \ref{sfig:ti_Tei_convergence}), which must be very low ($\sigma_\mu < 0.25~\mathrm{meV/atom} \approx 20~\mathrm{meV/supercell}$) to get an accurate $g_f$. Since $g_f$ is the difference between two large and similar numbers (the free energies of the bulk and defect supercells), small relative errors in either of these quantities can lead to a large error in $g_f$. The convergence tests resulted in the switching times reported in \ref{tab:t_switch}.

\begin{table}[ht]
\caption{Equilibration and switching times in picoseconds used for the thermodynamic integration paths of each system involved in the defect formation reaction. The timestep was set to \SI{2}{fs}.}\label{tab:t_switch}
\setlength{\tabcolsep}{8pt} 
\vspace{10pt}
\begin{tabular}{lcccc}
\hline
Path                              & CdTe & $\mathrm{Te_i^{+1}}$ & \kvc{V}{Te}{+2} & Te \\
\hline
Einstein $\rightarrow$ Anharmonic & 25, 100 & 25, 200 & 15, 150    &  25, 70   \\
100 K $\rightarrow$ 840 K         & 25, 170 & 25, 1000 & 25, 1000 &  25, 70   \\
840 K $\rightarrow$  500 K        &   -  &        -     &    -  &  25, 70   \\ 
\hline
\end{tabular}
\end{table}
We note that during the temperature scaling runs of the interstitial, defect diffusion occurs within the simulation timescale. This migration, which arises from the shape of the potential energy surface, contributes to the anharmonic free energy through the sampling of intermediate structures during site hopping. We expect that their contribution is small as the defects spend more time around their local minima configurations.

Finally, calculating the temperature dependence of the free energy for tellurium is slightly more challenging since it melts at \SI{722}{K}\cite{kracek_melting_1941}. Accordingly, we performed two TI simulations\cite{calphy}: i) Einstein crystal (\SI{100}{K}) $\rightarrow$ Anharmonic crystal (\SI{100}{K}) $\rightarrow$ Anharmonic crystal (\SI{840}{K}) and ii) Uhlenbeck-Ford model (\SI{840}{K}) $\rightarrow$ Liquid Te (\SI{840}{K}) $\rightarrow$ Liquid Te (\SI{500}{K}). By comparing the free energies from both simulations, we determined the phase transition temperature and the variation of the free energy with temperature (\ref{sfig:g_Te}). The calculated phase transition temperature is \SI{704}{K}, which is in good agreement with the experimentally reported value of \SI{722}{K}\cite{kracek_melting_1941}.

\emph{\textbf{Molecular dynamics.}}
To model the behaviour of the defects at room temperature, we performed NPT molecular dynamics with \texttt{LAMMPS}\cite{LAMMPS} using both a 65-atom ($a=13$~\AA) and 513-atom cubic supercells ($a=26$~\AA) to properly capture the dynamics and diffusion of the interstitial. The Nos\'e-Hoover thermostat and barostat were used (1 atm, 300~K and timestep of \SI{2}{fs}) with equilibration and production times of \SI{300}{ps} and \SI{1}{ns}, respectively. These trajectories were analysed with Python using tools from the \texttt{ase}\cite{ase-paper}, \texttt{pymatgen-analysis-defects}\cite{pymatgen_analysis_defects,shen2024simulating}, \texttt{pymatgen}\cite{Ong2013,Jain2013,Ong2015}, \texttt{dscribe}\cite{dscribe,dscribe2}, \texttt{umap}\cite{mcinnes_umap_2018}, \texttt{direct}\cite{qi_robust_2024},
\texttt{matplotlib}\cite{Hunter:2007}, 
and \texttt{seaborn}\cite{Waskom2021} packages, and visualised with \texttt{Ovito}\cite{ovito} and \texttt{CrystalMaker}\cite{crystalmaker}.\\
The energy barriers for the changes in configuration, orientation and position of $\mathrm{Te_i^{+1}}$ were calculated with the Nudge Elastic Band method\cite{henkelman_climbing_2000,henkelman_improved_2000}, as implemented in \texttt{ase}.\cite{ase-paper} The associated rates for these processes were calculated using Transition State Theory (e.g. $k(T)=\nu \exp{(-E_b/(k_BT))}$ and approximating the attempt frequency $\nu$ by the curvature of the PES at the initial state. 
The anharmonicity scores were calculated with \texttt{FHI-vibes}\cite{Knoop2020,knoop_anharmonicity_2020} on MD trajectories (NPT ensemble; 1~atm, 500~ps) at three temperatures (300, 550 and 900~K) for pristine CdTe and $\mathrm{Te_i^{+1}}$.

\section*{Declarations}
{\bf Data availability.}
The datasets and trained models will be available from a Zenodo repository upon publication.

{\bf Acknowledgments.}
We thank Se\'an R. Kavanagh for suggesting $\mathrm{Te_i^{+1}}$ as a test system and discussions regarding metastability. We also thank Venkat Kamil, Maurice de Koning, Talid Sinno, Blazej Grabowski, Sarath Menon and Luciano Colombo for useful discussions about thermodynamic integration. 
I.M.-L. thanks Imperial College London (ICL) for funding a President's PhD scholarship. 
J.K. acknowledges support from the Swedish Research Council (VR) program 2021-00486.
We are grateful to the UK Materials
and Molecular Modelling Hub for computational resources, which is partially funded by EPSRC (EP/P020194/1 and EP/T022213/1). This work used the ARCHER2 UK National
Supercomputing Service (https://www.archer2.ac.uk) via our membership of the UK’s HEC Materials Chemistry Consortium, funded by EPSRC (EP/L000202). We acknowledge the ICL High Performance Computing services for computational resources. 

{\bf Author contributions.}
Conceptualisation \& Project Administration: A.W., I.M.-L. Investigation, 
 methodology and formal analysis: I.M.-L. Supervision: A.W. Writing - original draft: I.M.-L. Writing - review \& editing: All authors. Resources and funding acquisition: A.W. These author contributions are defined according to the CRediT contributor roles taxonomy.

{\bf Competing interests.}
The authors declare no competing interests.

\bibliographystyle{rsc}
\bibliography{biblio}

\clearpage
\begin{center}
\textbf{Supplementary Information for `Point defect formation at finite temperatures with machine learning force fields'}
\end{center}

\renewcommand{\figurename}{Supplementary Figure}
\renewcommand{\tablename}{Supplementary Table}
\renewcommand\thefigure{S\arabic{figure}}    
\setcounter{figure}{0}
\renewcommand\thetable{S\arabic{table}}  
\setcounter{table}{0}
\renewcommand\thesection{\arabic{section}}  
\setcounter{section}{0}

\section{Machine learning force fields}
\subsection{Validation on test set}

\begin{table}[!ht]
\caption{
    Number of configurations in the datasets used to train and test each model. The training datasets are divided into training and validation sets (90\% and 10\%, respectively), with the latter used to monitor the validation loss during training. We note that we used more training configurations than necessary, as discussed in the \ref{subsec:learn_curve}.
}\label{tab:num_configs}
\vspace{10pt}
\begin{tabular}{ccccc}
\hline
& CdTe & $\mathrm{Te_i^{+1}}$ & \kvc{V}{Te}{+2} & Te \\
\hline
Train & 1412 & 3992 & 4316 & 1171 \\
Test & 300 & 237 & 132 & 312 \\
\hline 
\end{tabular}
\end{table}

\begin{figure}[ht]
    \centering
    \includegraphics[width=0.995\linewidth]{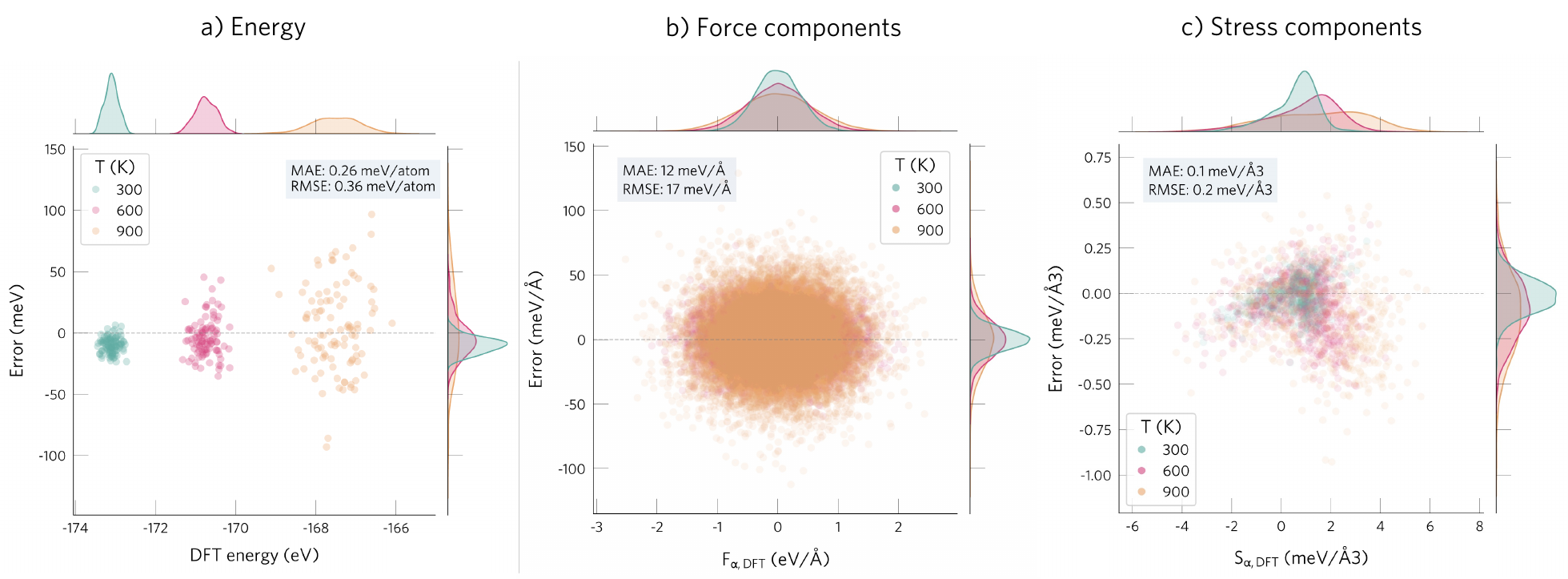}
    \caption{Distribution of mean absolute and root mean squared errors (MAE, RMSE) for the test set of bulk CdTe.}
    \label{sfig:errors_CdTe}
\end{figure}

\begin{figure}[ht]
    \centering
    \includegraphics[width=0.995\linewidth]{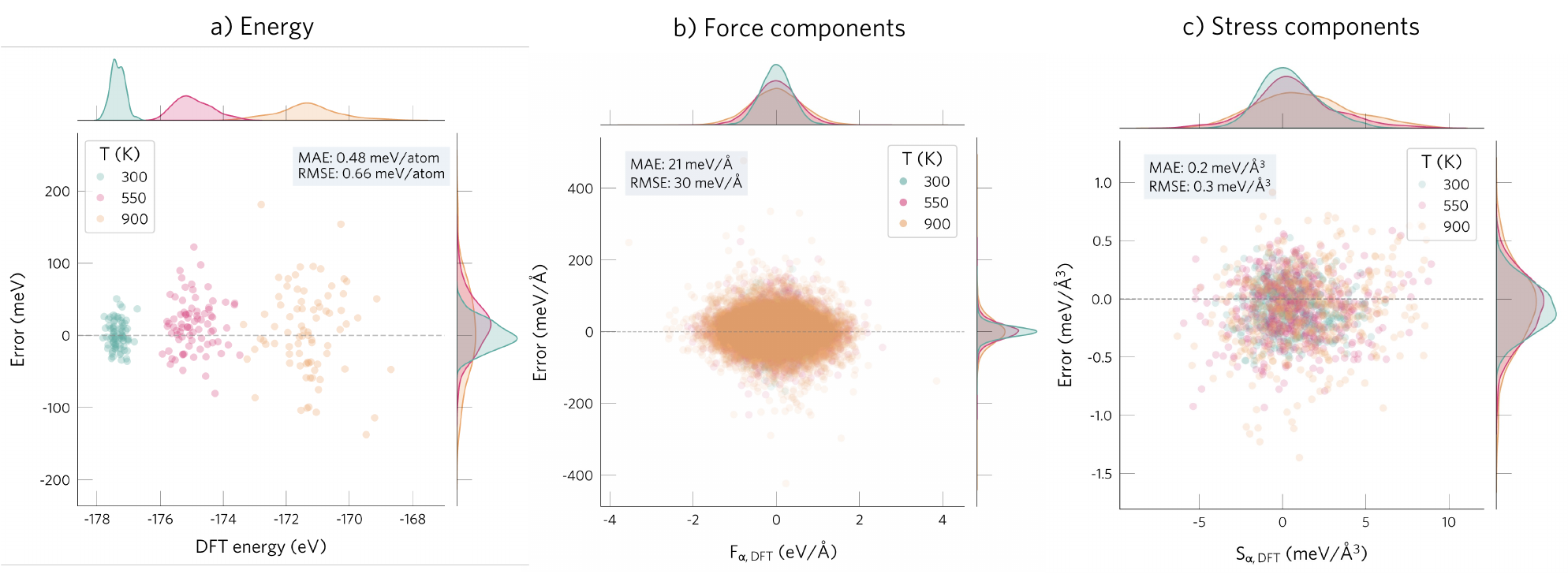}
    \caption{Distribution of mean absolute and root mean squared errors (MAE, RMSE) for the test set of $\mathrm{Te_i^{+1}}$. The distribution of the test and training configurations are illustrated in \ref{sfig:dataset_Tei}.}
    \label{sfig:errors_Tei}
\end{figure}

\begin{figure}[ht]
    \centering
    \includegraphics[width=0.995\linewidth]{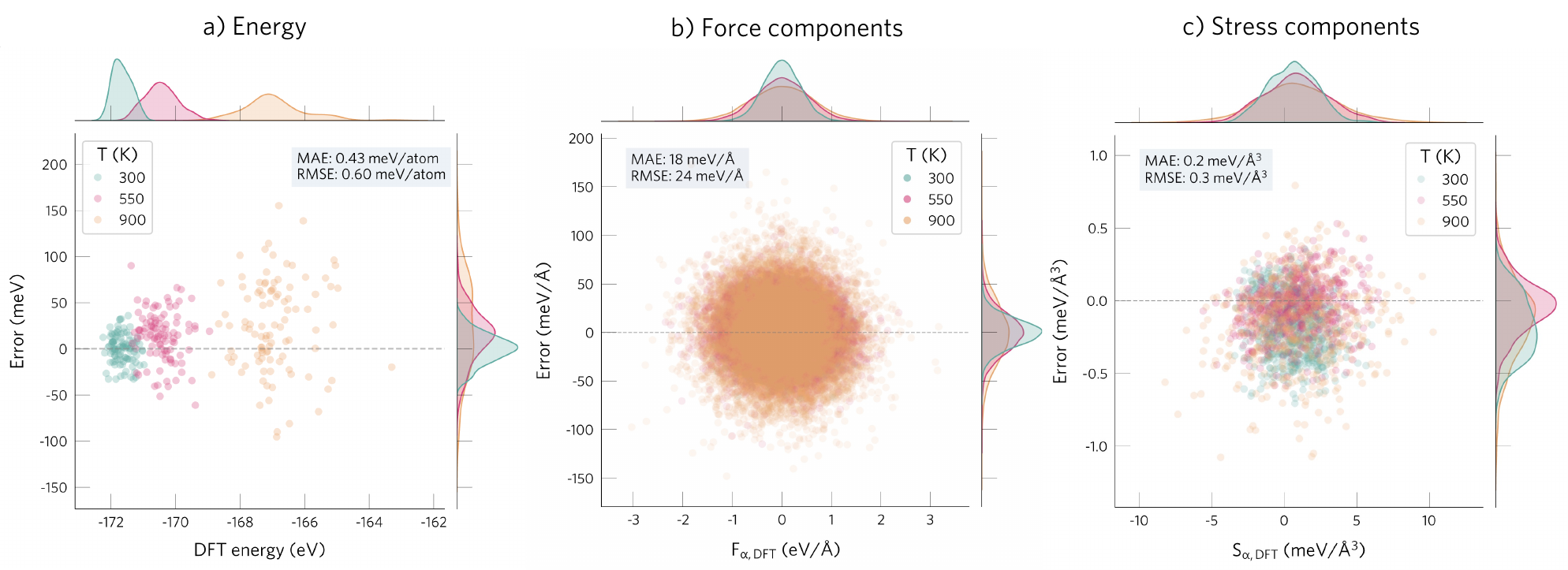}
    \caption{Distribution of mean absolute and root mean squared errors (MAE, RMSE) for the test set of\kvc{V}{Te}{2+}.}
    \label{sfig:errors_VTe}
\end{figure}

\begin{figure}[ht]
    \centering
    \includegraphics[width=0.995\linewidth]{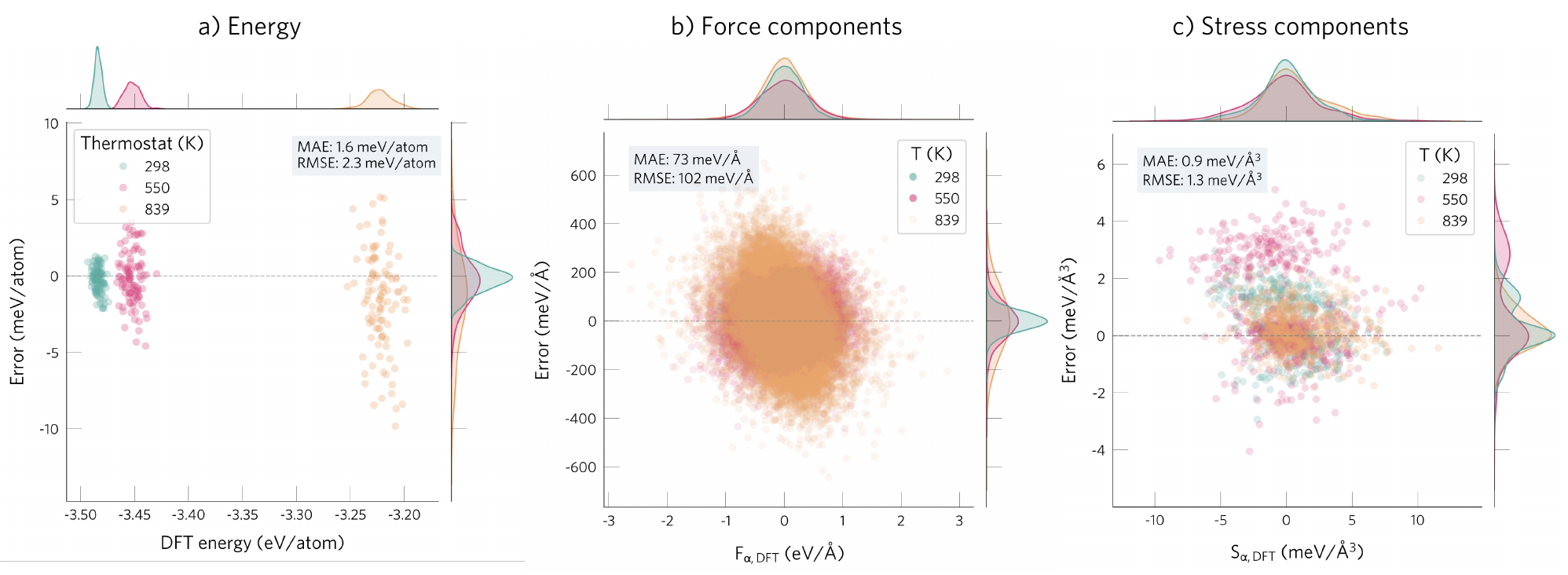}
    \caption{Distribution of mean absolute and root mean squared errors (MAE, RMSE) for the test set of Te. Note that Te melts at \SI{720}{\kelvin}, leading to larger errors for the liquid phase (T=\SI{900}{K}).}
    \label{sfig:errors_Te}
\end{figure}

\begin{figure}[ht]
    \centering
    \includegraphics[width=0.5\linewidth]{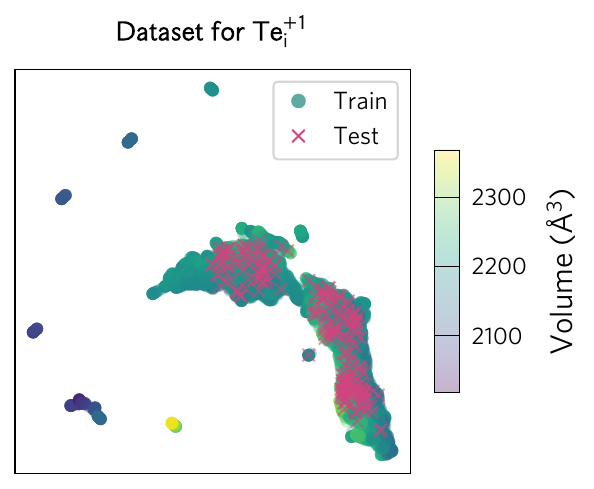}
    \caption{Two-dimensional feature map for the configurations in the dataset of $\mathrm{Te_i^{+1}}$. The configurations used for training are shown with circles, while the test ones are displayed with pink crosses. The isolated clusters of training datapoints correspond to compressed and expanded structures generated by scaling the equilibrium volume, which was necessary to ensure that the model could be applied with the quasiharmonic approximation. These regions were not included in the test set, which was designed to measure the accuracy of the model in typical application conditions (\SI{1}{atm}, 100-900~K), and accordingly expands over the configurations with equilibrium volumes at \SI{1}{atm}. The good coverage of the test set over the training set feature space demonstrates its ability to quantify the accuracy of the model. 
    Each configuration was encoded with its \texttt{DIRECT} descriptor (averaged over sites) and the dimensions were reduced using the Uniform Manifold Approximation, as implemented in the \texttt{UMAP} package\cite{mcinnes_umap_2018}.
    }
    \label{sfig:dataset_Tei}
\end{figure}

\begin{figure}[ht!]
    \centering
    \includegraphics[width=0.980\linewidth]{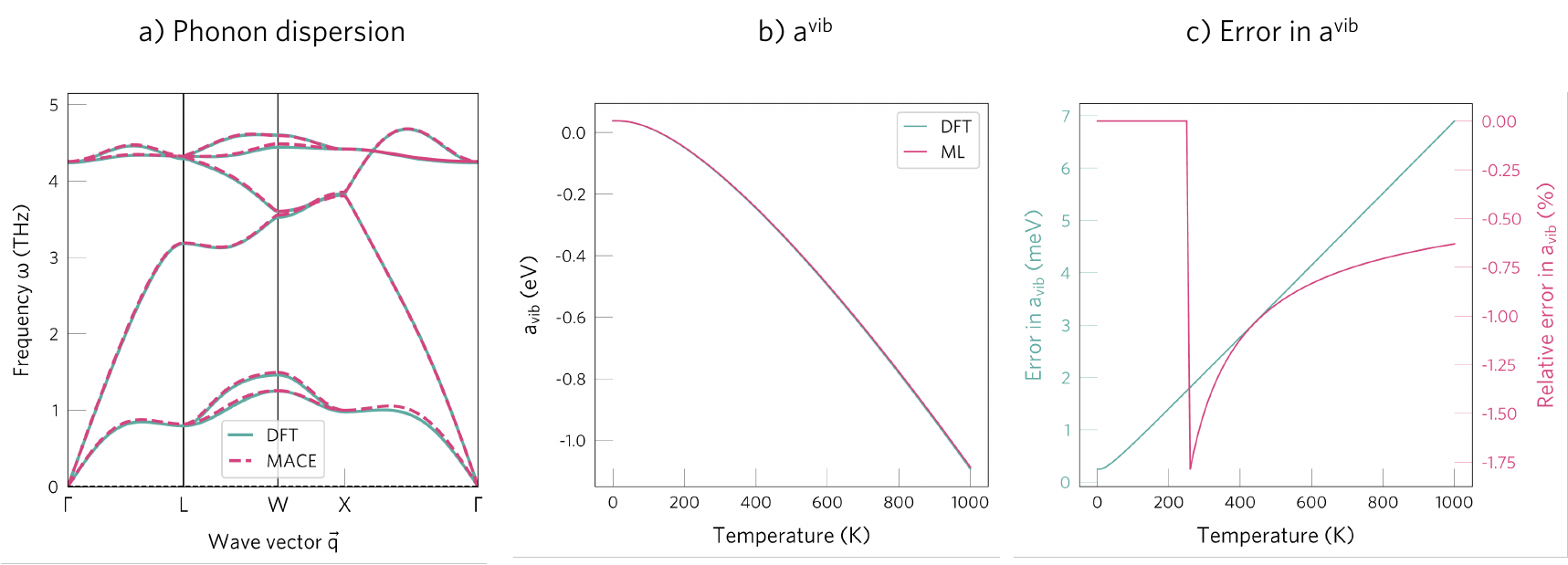}
    \caption{Comparison of the harmonic phonon dispersion and vibrational free energy calculated with DFT and the \texttt{DIRECT} MLFF for bulk CdTe (primitive cell, 2 atoms). Note that the abrupt change in the relative error is caused by $\mathrm{a^{vib}}$ changing sign (i.e., division by 0 in $\Delta a = (a^{ML} - a^{DFT})/a^{DFT}$).}
    \label{sfig:CdTe_phonon_dispersion}
\end{figure}
\FloatBarrier

\subsection{Learning curve}\label{subsec:learn_curve}
\begin{figure}[ht!]
    \centering
    \includegraphics[width=0.6\linewidth]{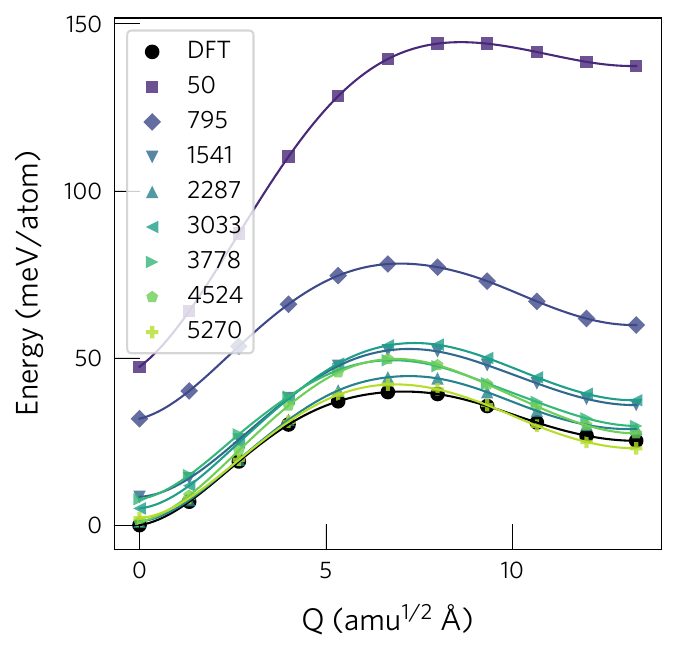}
    \caption{
    Performance of the models with increasing training sizes on the energy barrier between the stable structures. Even for models with a low test error ($\mathrm{MAE<0.5~meV/atom}$), the errors in the barrier are significant. This is caused by i) the small energy differences between the configurations ($\mathrm{E_{barrier,DFT}\approx40~meV/supercell=0.6~meV/atom}$, $\mathrm{\Delta E(C_{2v}-C_s)_{DFT}=18~meV/supercell=0.3~meV/atom}$) and ii) the fact that the configurations in the training set are sampled to maximise diversity. This method samples most structures from the high energy region of the PES (and thus less structures from the \SI{0}{K} path between the defect configurations). The legend denotes the number of configurations in the training set of each model, with the colormap ranging from dark purple to light green for increasing number of configurations. All energies are referenced to the DFT energy of the ground state configuration ($Q=0~\mathrm{amu}^{1/2}\angstrom$). 
    }
    \label{sfig:learning_curve_barrier}
\end{figure}

We analysed the learning curve for a model describing the behaviour of $\mathrm{Te_i^{+1}}$ at temperatures 100-900~K. We generated training sets with increasing number of configurations using the \texttt{DIRECT} method to sample the most diverse structures from the full dataset. This resulted in eight sets containing 50, 795, 1541, 2287, 3033, 3778, 4524 and 5270 configurations, which were used to train eight separate \texttt{DIRECT} models. By evaluating the performance of these models on the same test set (300 configurations, shown in \ref{sfig:dataset_Tei}), we can see that good accuracies ($\mathrm{MAE_E}\le1$~meV/atom) can be achieved with only 50 configurations (\ref{sfig:learning_curve_Tei}), as long as these are sampled to maximise their diversity. However, this accuracy level is not enough to properly capture the small energy differences between the stable configurations of $\mathrm{Te_i^{+1}}$, which only differ by $\mathrm{\Delta E(C_{2v}-C_s)_{DFT}=18~meV/supercell=0.3~meV/atom}$. As demonstrated in \ref{sfig:learning_curve_barrier}, training sets with at least 1500 configurations are needed to accurately describe the barrier. 

\begin{figure}[ht!]
    \centering
    \includegraphics[width=0.8\linewidth]{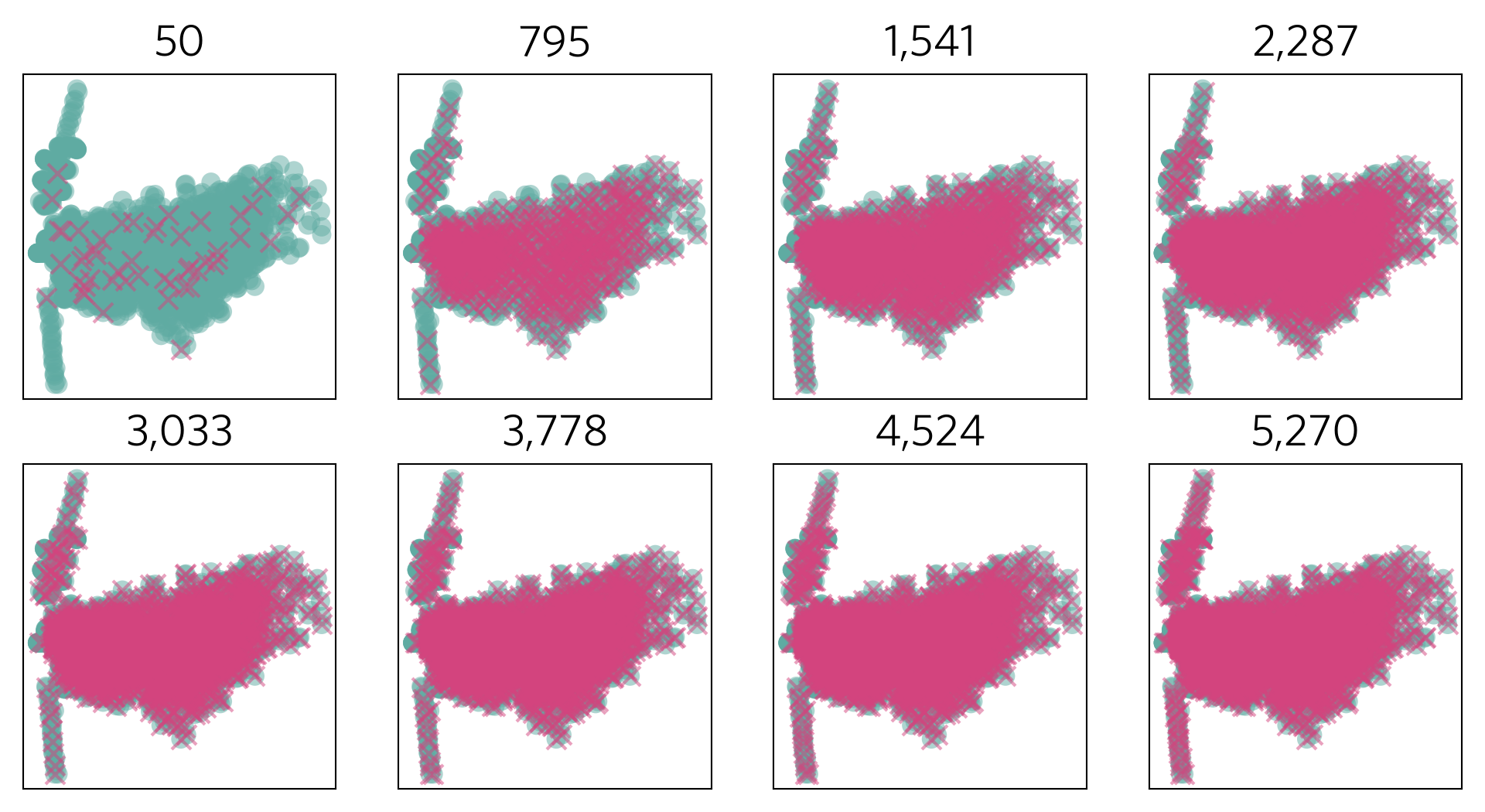} 
    \caption{
    Training sets of increasing size generated by sampling increasing number of configurations from the full dataset of 6016 $\mathrm{Te_i^{+1}}$ structures. Green circles illustrate the full dataset of configurations, while the selected structures are shown in pink crosses. Sampling was performed with the \texttt{DIRECT} algorithm\cite{qi_robust_2024}. Each structure was encoded with its \texttt{MACE} descriptor (averaged over sites) and the dimensions were reduced using Principal Component Analsysis, as implemented in the \texttt{maml} package\cite{maml}. 
    }
    \label{sfig:splits_sampling_Tei}
\end{figure}

\begin{figure}[ht!]
    \centering
    \includegraphics[width=0.6\linewidth]{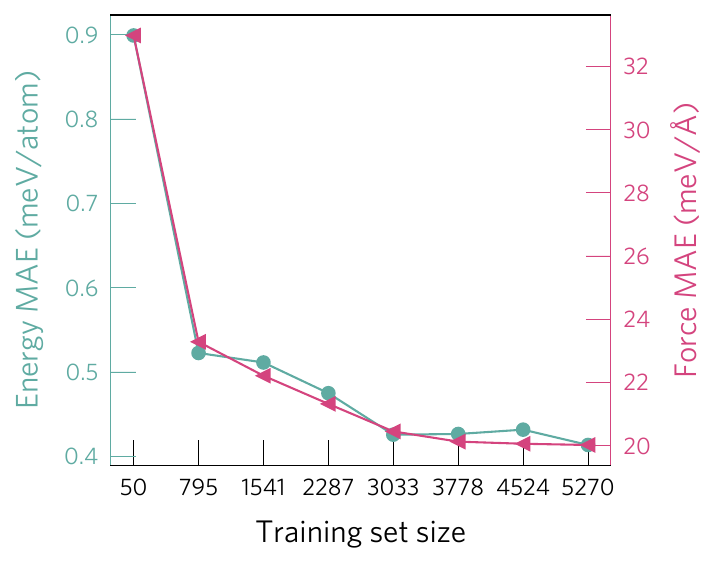}
    \caption{
    Learning curve for the $\mathrm{Te_i^{+1}}$ machine learning force field, showing the energy and force mean absolute errors (MAE) on the test set for models with an increasing number of training configurations. The root mean square error shows a similar trend to the MAE. Models of acceptable accuracy ($\mathrm{MAE_E}\le1$~meV/atom) can already be achieved with a small training set (50-1000 configurations), which is notable considering that the test set encompasses configurations from temperatures 300-900~K. 
    The different training sets were generated by sampling the full dataset using the \texttt{DIRECT} algorithm\cite{qi_robust_2024} to ensure optimal coverage of the configurational landscape, as illustrated in \ref{sfig:splits_sampling_Tei}. 
    }
    \label{sfig:learning_curve_Tei}
\end{figure}

\FloatBarrier
\subsection{Training machine learning force fields for defects}\label{subsec:mlff_defects}
\begin{figure}[ht!]
    \centering
    \includegraphics[width=1.0\linewidth]{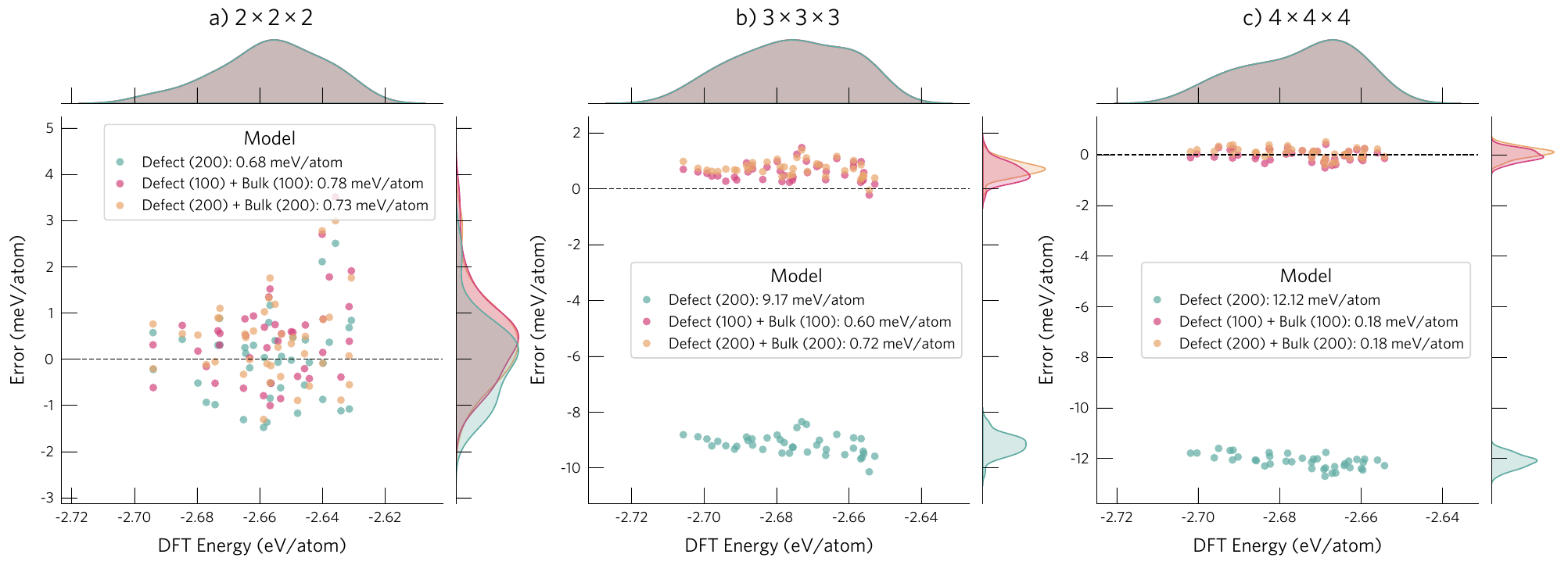}
    \caption{Comparison of test errors between the energies predicted by the $\mathrm{Te_i}^0$ models and DFT, when validated on supercells of different sizes: a) $2\times2\times2$, b) $3\times3\times3$ and c) $4\times4\times4$. Different colours correspond to the MACE models trained on different datasets: i) only defective structures of $\mathrm{Te_i}^0$ (200 configurations), ii) 100 $\mathrm{Te_i}^0$ structures and 100 pristine structures and iii) 200 $\mathrm{Te_i}^0$ structures and 200 pristine structures. For each supercell size, the 40 test structures used for validation were sampled from heating runs (from 300 to 600 K with a heating rate of \SI{0.7}{K/ps}) using the \texttt{DIRECT} method.
    }
    \label{sfig:mlff_supercell_size}
\end{figure}

In this study, we trained separate MACE force fields for the pristine and defective supercells since this lead to higher accuracies. However, we have observed that this approach limits the defect models when applied to \emph{larger supercells} than the ones used for training. While the models correctly describe the relative energies of the different configurations, they result in a systematic energy error. This is caused by how the MLFFs decompose the total energy into atomic contributions. When a model is trained only on \emph{defect} configurations, part of the energy associated with the formation of the defect (e.g. one additional/missing atom and broken/reformed bonds) is spread over the energies of all atoms in the supercell. If the model is then applied to a larger supercell, the predicted atomic energies for bulk-like atoms do not correspond to the energy of an atom in a bulk-like environment, leading to a systematic energy shift (\ref{sfig:mlff_supercell_size}.b). 

To solve this issue, one should train on \emph{both} defect and pristine supercells. To illustrate this point, we trained three MACE models on three different datasets: i) 200 defect configurations of $\mathrm{Te_i}$, ii) 100 defect ($\mathrm{Te_i}$) and 100 pristine configurations and iii) 200 defect ($\mathrm{Te_i}$) and 200 pristine configurations. Each dataset was sampled from the original training set ($2\times2\times2$ supercells) using the \texttt{DIRECT} method to maximise structural diversity. As illustrated in \ref{sfig:mlff_supercell_size}, while the model trained only on defect configurations leads to a smaller MAE when validating on configurations with the same number of atoms as the training structures ($2\times2\times2$ supercell, 65 atoms), it results in a systematic error when applied to larger supercells ($3\times3\times3$ (217 atoms) and $4\times4\times4$ (513 atoms)). This leads to a general conclusion when training MLFFs for point or extended defects: the training dataset should include both defect and pristine configurations if the models will be applied to larger defect supercells.

\newpage
\FloatBarrier
\section{Dynamic analysis}

\begin{figure}[!ht]
    \centering
    \includegraphics[width=0.5\linewidth]{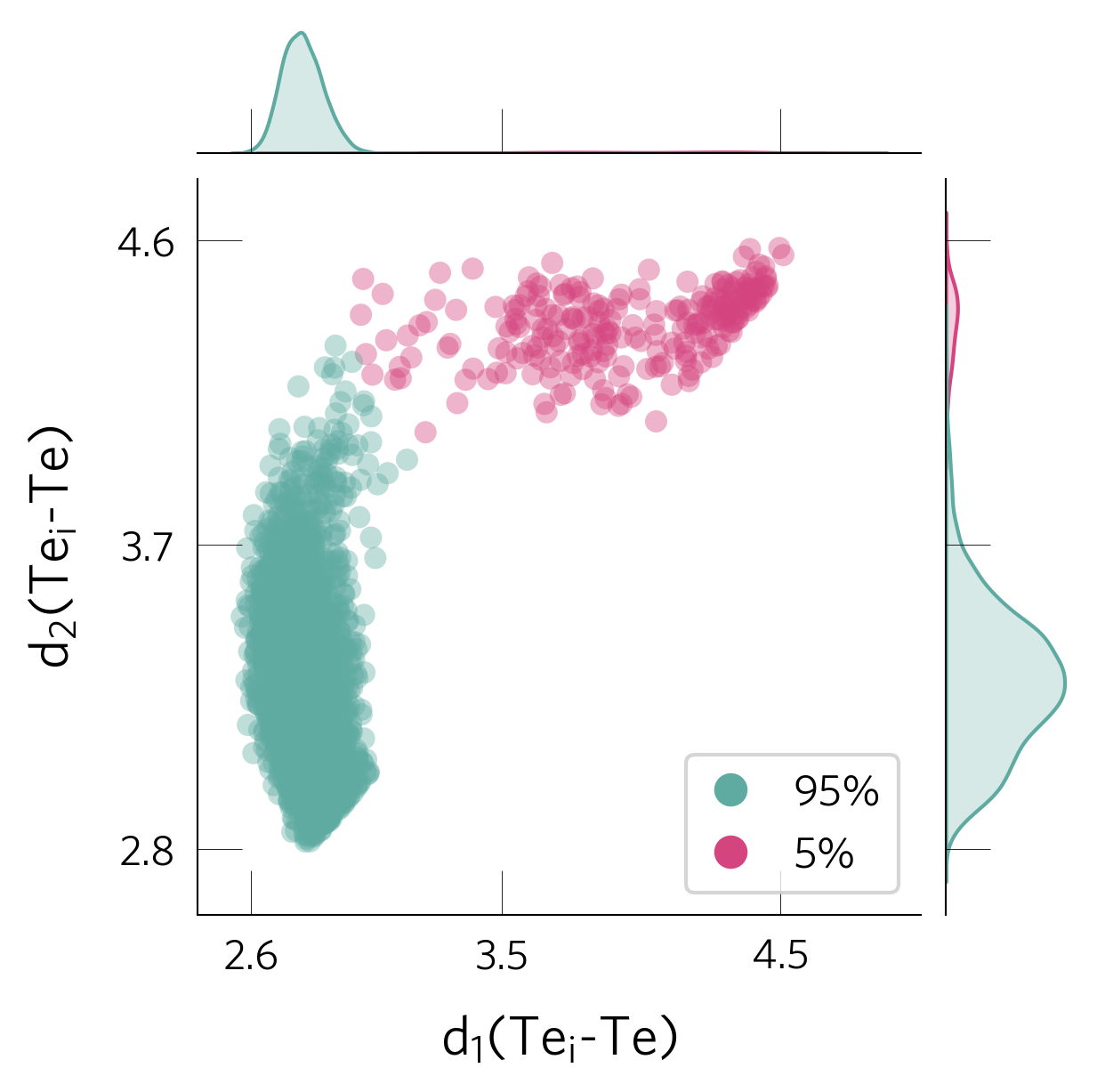}
    \caption{Changes in the configuration of $\mathrm{Te_i^{+1}}$ at 300~K, quantified by monitoring the shortest Te$\mathrm{_i}$-Te distances. While there is little variation in the shortest Te-Te bond (localised peak for $d_1$), the second shortest bond shows a wide variation ($d_2=2.8-4.0$~\angstrom) --- illustrating the change between the low-energy metastable configurations. Finally, there are some configurations where the interstitial occupies a significantly less favourable position without any Te-Te bonds (pink data points).}
    \label{sfig:Tei_distance_map}
\end{figure}

\begin{figure}[!ht]
    \centering
    \includegraphics[width=0.9\linewidth]{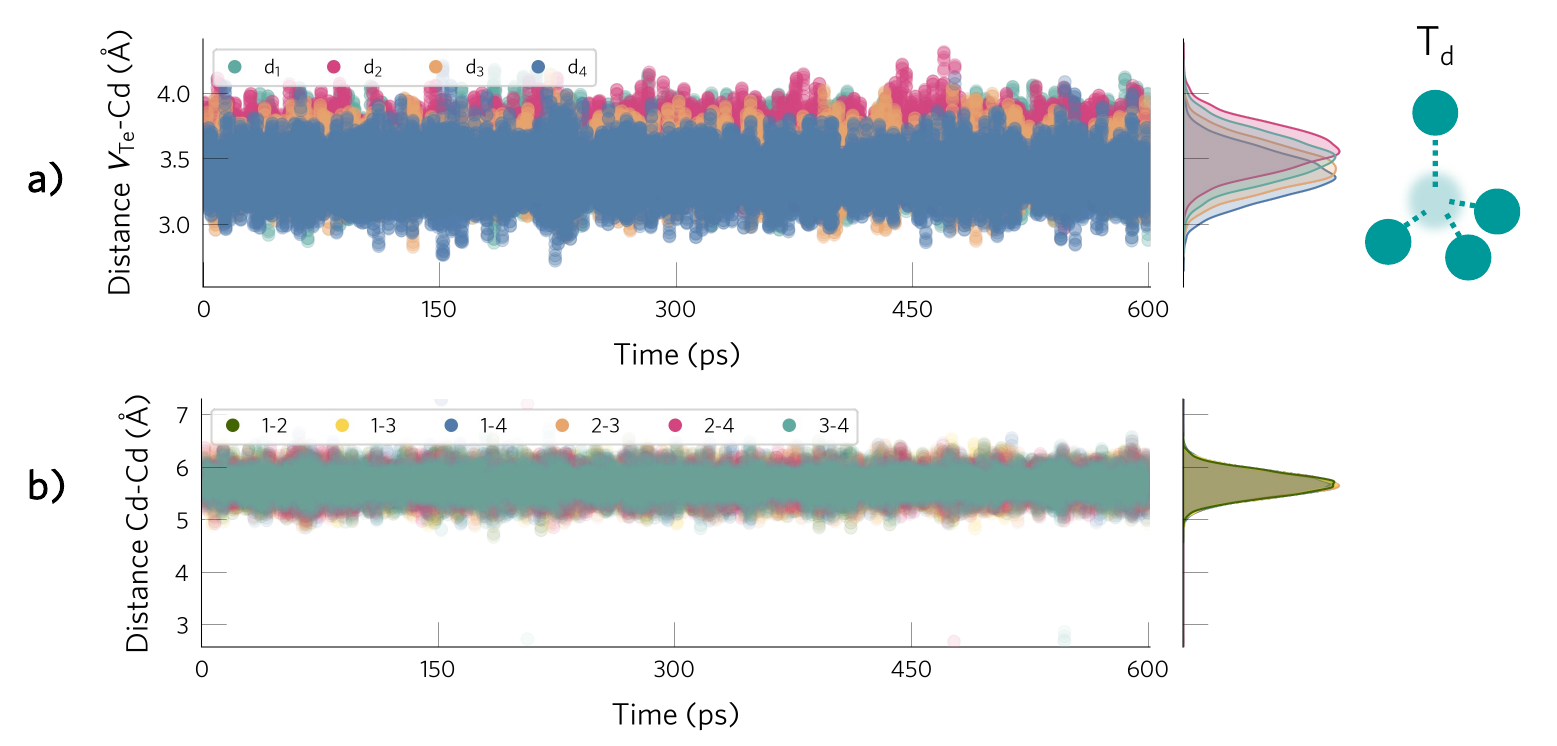}
    \caption{Evolution of \kvc{V}{Te}{+2} at 1 atm and 300~K (NPT ensemble), which stays in its $T_d$ configuration. a) Distance between the vacancy and its Cd neighbours, showing that all of them stay at a similar distance. b) Distance between the four Cd atoms neighbouring the vacancy, with these distances being similar and showing little variation with time.}
    \label{sfig:dynamics_VTe}
\end{figure}

\begin{figure}[!ht]
    \centering
    \includegraphics[width=0.85\linewidth]{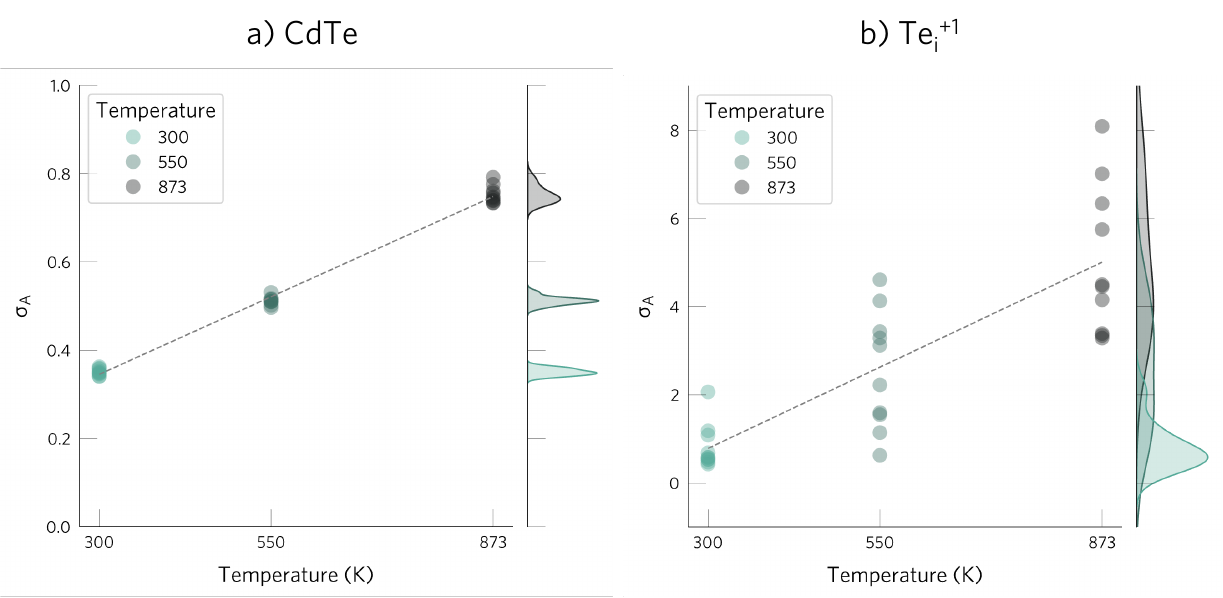}
    \caption{Anharmonicity scores\cite{knoop_anharmonicity_2020} for bulk CdTe and $\mathrm{Te_i^{+1}}$, calculated on ten independent NPT trajectories (1~atm, T$=$300, 550, 873~K). The high anharmonicity scores of $\mathrm{Te_i^{+1}}$ are caused by its dynamic character (changes in configuration and position). For both bulk and the defect, the anharmonic character increases with temperature as the vibrational amplitude increases. 
    }
    \label{sfig:anh_scores}
\end{figure}

\FloatBarrier
\section{Free energies}

\begin{figure}[!ht]
    \centering
    \includegraphics[width=0.45\linewidth]{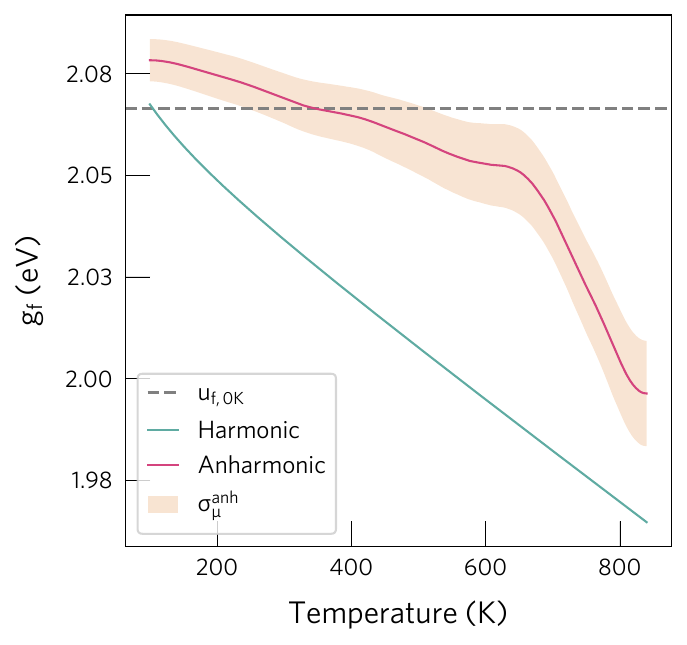}
    \caption{Comparison of approximations for calculating the defect formation free energy, $g_f(T)$, of \kvc{V}{Te}{+2}. For comparison, the formation internal energy, $u_f(0~\mathrm{K}$), typically used in defect studies, is shown with a dashed grey line. This comparison shows that for \kvc{V}{Te}{+2}, entropic effects are negligible and barely affect $u_f(0~\mathrm{K})$ --- which agrees with its static behaviour at room temperature (\ref{sfig:dynamics_VTe}). Note that the change in slope of $g_f^{anh}$ is caused by the change in the free energy of Te, which melts at \SI{704}{K}.}
    \label{sfig:g_f_V_Te}
\end{figure}

\begin{figure}[!ht]
    \centering
    \includegraphics[width=0.5\linewidth]{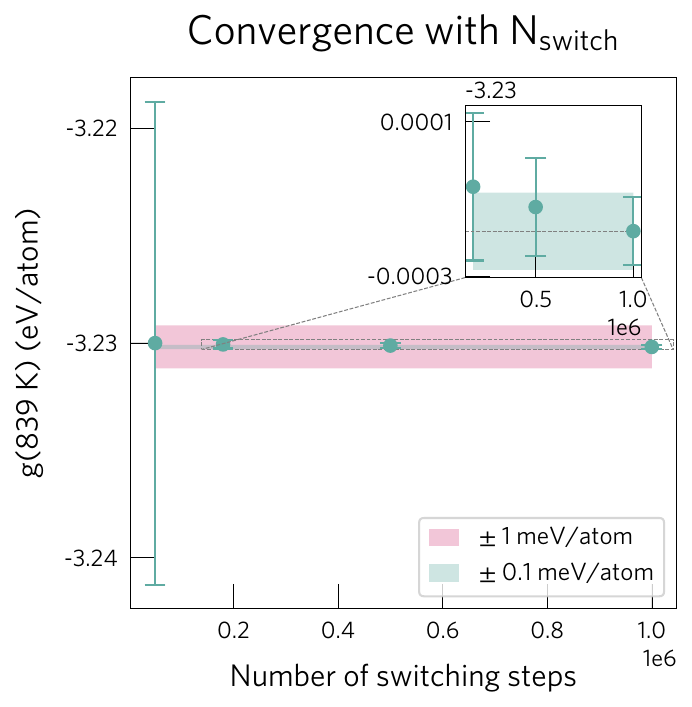}
    \caption{Convergence of the temperature-dependent free energy of $\mathrm{Te_i}$ with respect to the number of switching steps used in the non-equilibrium thermodynamic integration simulations. The convergence is evaluated by comparing the values of the free energy at the end temperature (\SI{839}{K}). Note that differences below \SI{0.1}{meV/atom} are achieved for $N_\mathrm{switch} \geq 1.8 \times 10^5$ steps. The timestep was set to \SI{2}{fs} and the uncertainties are determined by the mean standard error between 10 independent simulations performed for each value of $N_\mathrm{switch}$.}
    \label{sfig:ti_Tei_convergence}
\end{figure}

\begin{figure}[!ht]
    \centering
    \includegraphics[width=0.45\linewidth]{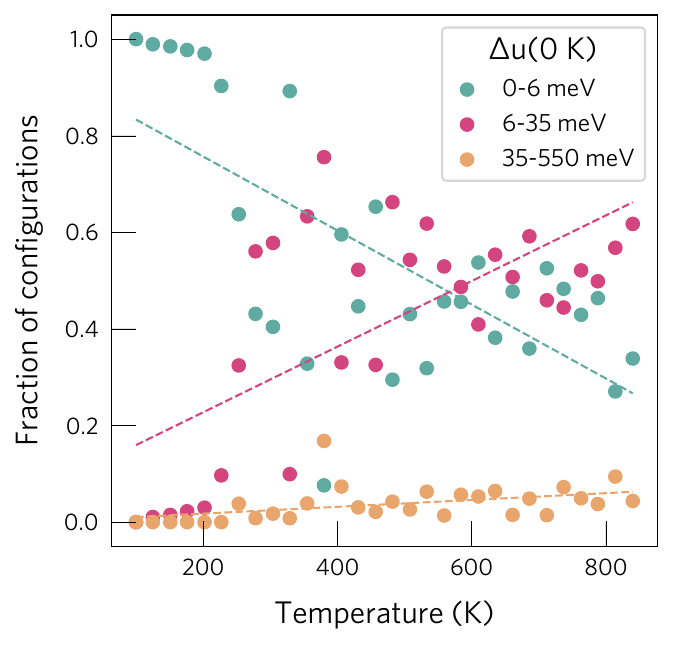}
    \caption{Distribution of $\mathrm{Te_i^{+1}}$ in the ground state and metastable structures at different temperatures. The populations of the configurations are determined using the `Inherent structures' formalism, by performing NPT MD simulations at different temperatures (\SI{80}{ps}, \SI{1}{atm}), sampling 1600 equally-spaced configurations and relaxing them to their  \SI{0}{K} local minima using a conjugate gradient optimiser\cite{luo_configurational_2022}. While at low temperatures (T $>$ \SI{200}{K}), the defect resides only in its lowest energy structure, above this temperature the population of the metastable configuration (i.e. the structure with higher internal energy) increases until it reaches the value of the ground state structure. 
    }
    \label{sfig:ratio_configs_Tei}
\end{figure}

\begin{figure}[!ht]
    \centering
    \includegraphics[width=0.45\linewidth]{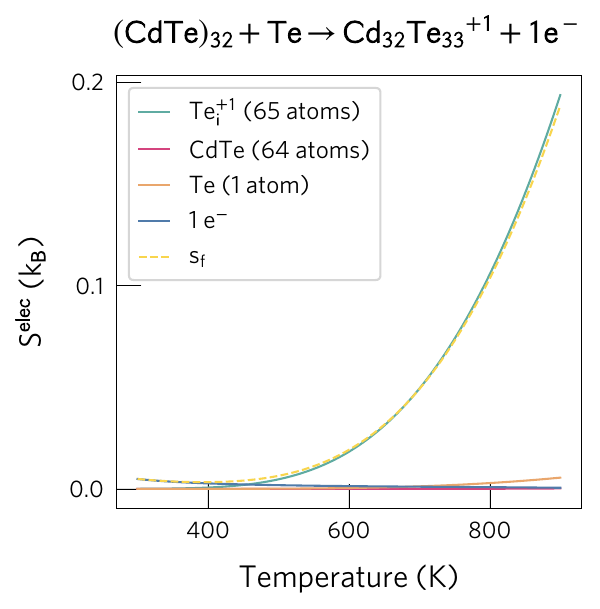}
    \caption{Electronic entropy for the formation of $\mathrm{Te_i^{+}}$. The increase in entropy is caused by the defect introducing an empty electronic level 0.7 eV above the valence band maximum.
    }
    \label{sfig:s_elec}
\end{figure}

\begin{figure}[!ht]
    \centering
    \includegraphics[width=0.45\linewidth]{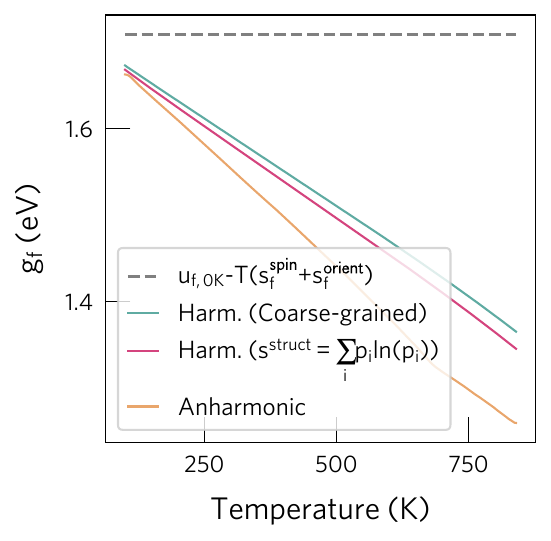}
    \caption{Comparison of methods to calculate the structural or configurational entropy for the formation of $\mathrm{Te_i^{+}}$. The coarse-grained approach uses Equation (16) from the main text while the second method uses Equation (11), as explained in the Methods. Both approaches result in a similar formation free energy.
    }
    \label{sfig:s_struc_coarse_grained}
\end{figure}

\begin{figure}[!ht]
    \centering
    \includegraphics[width=0.45\linewidth]{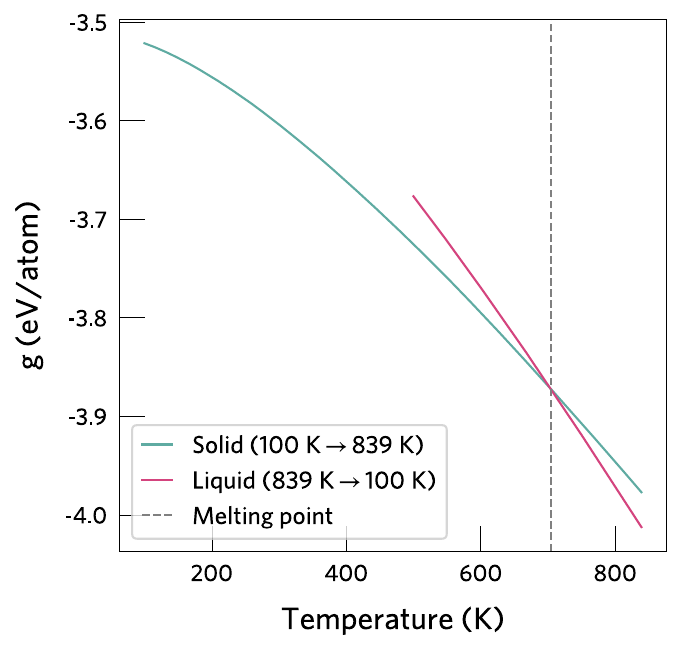}
    \caption{Temperature variation of the Te free energy calculated with thermodynamic integration, showing the two independent integration paths and the resulting melting point (\SI{704}{K}), in reasonable agreement with the previously reported value of \SI{722}{K}\cite{kracek_melting_1941}.
    }
    \label{sfig:g_Te}
\end{figure}
%

\end{document}